\documentstyle[amssymb,twocolumn,prl,aps,graphicx]{revtex}
\large

\begin{document}
\draft
\twocolumn[
\hsize\textwidth\columnwidth\hsize\csname @twocolumnfalse\endcsname

\title{Density Fluctuations in Molten Lithium: \\
Inelastic X-Ray Scattering Study. }
\author{
    T.~Scopigno$^{1}$,
    U.~Balucani$^{2}$,
    G.~Ruocco$^{3}$,
    F.~Sette$^{4}$}
\address{
    $^{1}$Dipartimento di Fisica and INFM, Universit\'a di Trento, I-38100,
Povo, Italy.\\
    $^{2}$Istituto di Elettronica Quantistica CNR, I-50127, Firenze, Italy.\\
    $^{3}$Dipartimento di Fisica and INFM, Universit\'a di L'Aquila, I-67100,
L'Aquila, Italy.\\
    $^{4}$European Synchrotron Radiation Facility, B.P. 220 F-38043 Grenoble,
Cedex France.
    }
\date{\today}
\maketitle
\begin{abstract}
New inelastic X-ray scattering experiments have been
performed in liquid lithium at two different temperatures: $T=475$ K
(slightly above the melting point) and $600$ K. Taking advantage of the
absence of any kinematical restriction and incoherent contribution, and
pushing the instrumental resolution up to $1.5$ meV, it was possible to
perform an accurate investigation of the dynamic structure factor $S(Q,E)$
in the wavevector range from $1$ to $110$ nm$^{-1}$. For $Q$ smaller than
$Q_m\simeq 25$ nm$^{-1}$, the position of the main peak of the static
structure factor, a detailed analysis of the lineshapes shows that any
picture of the relaxation mechanisms based on a simple viscoelastic model
must be abandoned. All the spectral features can instead satisfactorily be
accounted for by including both fast and slow relaxation processes. The
physical origin of the slow relaxation is associated to the structural
rearrangement, while the local nature of the fast one is extensively
discussed. At larger $Q$ values a gradual crossover from the strongly
correlated to single particle dynamics occurs, with an important weight
provided by quantum effects.
\end{abstract}
]


\section{INTRODUCTION}

The dynamics of liquid metals has been extensively investigated in the
recent past with the main purpose of ascertaining the role of the mechanisms
underlying both collective and single-particle motions at the microscopic
level. In the special case of collective density fluctuations, it is known
that well-defined oscillatory modes can be supported even outside the strict
hydrodynamic region. In molten alkali metals, this feature is found to
persist down to wavelengths of one or two interparticle distances, making
these systems excellent candidates to test the various theoretical
approaches developed so far for the microdynamics of the liquid state. As a
consequence, since the pioneering inelastic neutron scattering (INS) study
by Copley and Rowe \cite{rubidio2} in liquid rubidium the interest in
performing more and more accurate experiments is continuously renewed: INS
investigations have been devoted to liquid cesium \cite{cesio}, sodium \cite
{sodio}, lithium \cite{verkerk}, potassium \cite{potassio} and again
rubidium \cite{rubidio}.

Up to a few years ago the only experimental probes adequate to access the
interparticle distance region in collective dynamics were thermal neutrons.
With this technique foundamental results have been achieved in the field
of condensed matter structure and dynamics.
Unfortunately, in several systems their use for the determination of the
dynamic structure factor $S(Q,E)$ becomes extremely difficult (if not
impossible) for two reasons. The first one reflects the presence of an
incoherent contribution to the total neutron scattering cross section. In
liquid sodium, for example, the incoherent cross section dominates; even in
more favorable cases (Li, K) at small $Q$ the intensity of the collective
contribution is low, and its extraction requires a detailed knowledge of the
single particle dynamics. The second reason is dictated by the need of
satisfying both the energy and momentum conservation laws which define the $%
(Q-E)$ region accessible to the probe \cite{libroumberto}. Roughly speaking,
when the sound speed of the system exceeds the velocity of the probing
neutrons ($\sim 1500$ m/s for thermal neutrons) collective excitations
cannot be detected for $Q$ values below $Q_m,$ the position of the first
sharp diffraction peak of the sample, namely just in the most significant
region for the collective properties.

By virtue of the $m^{-1/2}$ dependence of the isothermal sound speed $c_0$
(see section \ref{def}) the higher the atomic number of the system, the
wider is the kinematic region accessible to neutrons, so that accurate INS
data are available for rubidium ($c_0\sim 1260$ m/s) \cite{rubidio,rubidio2}
and cesium ($c_0\sim 970$ m/s) \cite{cesio}, while more difficulties are met
in the case of lighter metals. In particular, lithium represents the most
critical case due to its high sound speed ($c_0\sim 4500$ m/s) and to the
equality of the coherent and incoherent cross sections: for this reason INS
aiming to the study of collective properties of Li represents a very hard
challenge \cite{tesidejong}. From a general point of view the main outcome
of all these INS experiments, as far as collective properties are concerned,
is the evidence of inelastic excitations in $S(Q,E)$ that, in the specific
cases of rubidium and cesium, exhibit a {\it positive dispersion, }i.e. an
increase of $\omega _m(Q)$ (the position of the $S(Q,\hbar \omega )$ peaks)
with respect to the values implied by the hydrodynamic sound speed. About
the details of the relaxations driving such dispersion not much can be
inferred at this level: the experimental results have been necessarily
analyzed within simple models such as the damped harmonic oscillator \cite
{rubidio} or the kinetic model \cite{jong,rubidio}, suitable to extract
reliable and resolution-corrected information on the peak positions but not
about the lineshape features. Some extra efforts have been done in the case
of cesium where informations about an average relaxation time have been
extracted \cite{cesio}.

A very useful tool, complementary to the ''traditional'' experimental
facilities is the numerical simulation technique, in particular Molecular
Dynamics (MD): the choice of a realistic interatomic potential - i.e. a
potential model able to reproduce structural properties - allows the
determination of the dynamics of the system via the integration of the
classical Newton equations. In this framework the single particle and the
collective properties can easily be investigated within technical
restrictions due to the finite box size (defining the minimum accessible
wavevector) and computation time (related to the statistical quality and to
the energy resolution of the calculated spectra). Broadly speaking the
features of the atomic collective motion i.e. the details of the $S(Q,E)$
lineshape, as outcome of MD run, turns out to be less noisy and more
straightforward than the correspondent INS results: no absolute
normalizations are required, no mixing between coherent/incoherent dynamics
occurs and, above all, basically no resolution corrections are needed.

After the first experiments on rubidium, a considerable number of MD works
have been published in this field, giving a valuable support to the
experimental measurements \cite{rahman1,MDworks}. On
the specific case of liquid lithium, the performed MD simulations \cite
{umblitiomd,canaleslitiomd} show features common to all the other molten
alkali metals, like the presence of {\it positive dispersion - }a
particularly relevant result that could have not be achieved with neutrons
for the previously mentioned reasons. At temperature around $T_m$ the
increase of the sound velocity is of the order of 20\%, and not much is
known about its microscopic origin.

In the recent past, the development of new synchrotron radiation facilities
opened the possibility of using X-rays to measure the $S(Q,\omega)$ (which
is proportional to the scattered intensity) in the non-hydrodynamic region;
in this case the photon speed is obviously much larger than the excitations
velocity, and no kinematic restriction occurs. Moreover, in a monatomic
system as lithium, the scattering cross section is purely coherent and so it
is directly associated with the dynamic structure factor. Some experiments
have been performed on liquid lithium \cite
{burk,burkel1,burkel2,sinn,sinnthesis} with progressively increasing
resolution; it has been possible to show the existence of propagating
collective excitations, but due to resolution limitations (never below $%
\Delta E_{FWHM}\sim 12$ meV), no detailed information about the lineshape
could be extracted.

In this work we report the results of Inelastic X-ray Scattering (IXS)
experiments performed on liquid lithium with very high energy resolution,
adequate to probe accurately the detailed features of $S(Q,E)$ - this allows
to investigate whether the lineshape dependence on momentum transfer can be
interpreted within the phenomenology of one or more relaxation processes. In
particular in Section II we review the basic theoretical framework adopted
for the interpretation of the IXS results. The latter are reported in
Section III together with a brief account of the experimental setup. Section
IV is devoted to the data analysis and discussion. The main outcome of the
present paper is finally summarized in Section V along with some concluding
remarks.

\section{BASIC THEORY\label{basic}}

\subsection{Definitions\label{def}}

In the classical limit, the basic time correlation probing collective
dynamics in a monatomic fluid ($N$ particles with mass $m$) is the
intermediate scattering function

\begin{equation}
F(Q,t)=(1/N)\sum_{i,j}\left\langle e^{-i{\bf Q}\cdot {\bf r}_i(0)}e^{i{\bf Q}%
\cdot {\bf r}_j(t)}\right\rangle   \label{fqt}
\end{equation}
where ${\bf r}_i(t)$ denotes the position of the {\it j}-th particle at time
$t$. The dynamic structure factor $S(Q,\omega )$ is the frequency spectrum
of $F(Q,t)$, while the structural features are accounted for the initial
value $F(Q,t=0)=S(Q)$. The quantity $F(Q,t)$ then obeys the equation

\begin{eqnarray}
\stackrel{..}{F}(Q,t)&&+\omega _0^2(Q)F(Q,t) \\ &&+\int_0^tM(Q,t-t^{\prime })%
\stackrel{.}{F}(Q,t^{\prime })dt^{\prime }=0  \label{langevin} \nonumber
\end{eqnarray}
where

\begin{equation}
\omega _0^2(Q)=K_BTQ^2/mS(Q)  \label{a2}
\end{equation}
is the second classical normalized frequency moment of $S(Q,\omega ),$ while
$M(Q,t)$ is the so called memory function of the system, related in some way
to the details of the Hamiltonian \cite{mori}. In the limit $Q\rightarrow 0,$
$\omega _0^2(Q)\rightarrow c_0^2Q^2$, where $c_0$ is the isothermal sound
velocity; hence the quantity $c_0$ $(Q)\equiv \omega _0(Q)/Q$ can be
interpreted as a suitable generalization of $c_0$ to finite wavevectors.

From Eq.(\ref{langevin}) a formally exact representation of the Laplace
transform of $F(Q,t)$ can be written as \cite{libroumberto,BY}

\begin{eqnarray}
\widetilde{F}(Q,z) &=&\int_0^\infty dte^{-zt}F(Q,t)  \label{fqz} \\
&=&S(Q)\left\{ {z+}\frac{{\omega }_0^2{(Q)}}{{[z+}\widetilde{M}{(Q,z)]}}%
\right\} ^{-1}  \nonumber
\end{eqnarray}

From the knowledge of $\widetilde{F}(Q,z)$ one straightforwardly obtains $%
S(Q,\omega )=(1/\pi )Re\widetilde{F}(Q,z=i\omega )$ in terms of the real ($%
M^{\prime }$) and imaginary ($M^{\prime \prime }$) parts of the
Fourier-Laplace transform of the memory function:
\begin{equation}
S(Q,\omega )=\frac{S(Q)}\pi \frac{\omega _0^2(Q)M^{\prime }(Q,\omega )}{%
\left[ \omega ^2-\omega _0^2+\omega M^{\prime \prime }(Q,\omega )\right]
^2+\left[ \omega M^{\prime }(Q,\omega )\right] ^2}  \label{sqwgenerale}
\end{equation}

The spectral features of the dynamic structure factor can be characterized
by its frequency moments $\Omega _S^{(n)}(Q)\equiv \int \omega ^nS(Q,\omega
)d\omega $, where, for a classical system, only the even frequency moments
(such as $\Omega _S^{(0)}(Q)=S(Q)$ and $\Omega _S^{(2)}(Q)=S(Q)\omega
_0^2(Q) $) are different from zero.

In the following, we shall also find convenient to consider the
''longitudinal current spectrum'' defined as $C_{L}(Q,\omega )=\left( \omega
^{2}/Q^{2}\right) S(Q,\omega )$. The presence of the factor $\omega ^{2}$
wipes out the low frequency portion of the dynamic structure factor, and
consequently emphasizes the genuine inelastic features of $S(Q,\omega )$.
After its definition and Eq.(\ref{fqz}), it is readily seen that the Laplace
transform $\widetilde{C}_{L}(Q,z)$ satisfies

\begin{eqnarray}
\widetilde{C}_L(Q,z) &=&-z[z\widetilde{F}(Q,z)-S(Q)]  \label{cqz} \\
\ &=&\left\{ {z+[\omega _0^2(Q)/\ z]\ +\ \widetilde{M}(Q,z)}\right\} ^{-1}
\nonumber
\end{eqnarray}
Again, the spectrum $C_L(Q,\omega )$ can be expressed as $(1/\pi )Re%
\widetilde{C}_L(Q,z=i\omega )$. Then the position and the width of the
inelastic peaks in $C_L(Q,\omega )$ are determined by the poles of $%
\widetilde{C}_L(Q,z)$.

The above definitions are valid for a classical system. In the case of our
interest, the main effect of quantum-mechanical corrections stems from the
well-known inequality of the positive and negative-frequency parts of the
spectra, connected by the detailed balance factor $e^{\beta \hbar \omega }$.
Additional sources of non-classical behavior, such as those associated with
a finite value of the deBoer wavelength $\Lambda =\left( 2\pi \hbar
^2/mK_BT\right) ^{1/2}$, are small (in the explored lithium states $\Lambda $%
\ is only $0.11$ times the average interparticle distance) and can safely be
neglected. Since the effects of the detailed balance are clearly visible in
the experimental IXS spectra, we briefly discuss a possible procedure to
account for this quantum feature in a consistent way, while preserving the
inherent advantages of the classical description. In doing this, for the
sake of clarity we shall denote all the previous classical quantities with
the subscript {\it cl}, while the notation {\it q} will refer to the quantum
case.

The natural theoretical counterpart of the classical density correlation
function is the so called Kubo canonical relaxation function \cite{kubo}

\begin{equation}
K_q(Q,t)=\frac 1{\beta N}\sum_{i,j}\int_0^\beta d\lambda \left\langle e^{-i%
{\bf Q\cdot }\widehat{{\bf r}}_i(0)}e^{-\lambda \widehat{H}}e^{i{\bf Q\cdot }%
\widehat{{\bf r}}_j(t)}e^{\lambda \widehat{H}}\right\rangle  \label{kubo}
\end{equation}
where $\beta =1/K_BT$ and the angular brackets denote a quantum statistical
average. In the classical limit ($\beta \rightarrow 0,$ $\hbar \rightarrow 0$%
) the operators $\widehat{A}$ become classical commuting dynamical variables
and $K_q(Q,t)\rightarrow F_{cl}(Q,t).$ It can be shown \cite{loves} that $%
K_q(Q,t)$ is a real even function of time, so that its spectrum $%
K_q(Q,\omega )$ is an even function of frequency. On the other hand, the
experimental scattering cross section involves the Fourier transform $%
S_q(Q,\omega )$ of the quantum density correlator $F_q(Q,t)=(1/N)\sum_{i,j}%
\left\langle e^{-i{\bf Q\cdot }\widehat{{\bf r}}_i(0)}e^{i{\bf Q\cdot }%
\widehat{{\bf r}}_j(t)}\right\rangle .$ The relation between $S_q(Q,\omega )$
and $K_q(Q,\omega )$ reads \cite{loves}

\[
S_{q}(Q,\omega )=\frac{\beta \hbar \omega }{1-e^{-\beta \hbar \omega }}%
K_{q}(Q,\omega )
\]
and satisfies the detailed balance condition. Moreover the condition

\begin{equation}
\Omega _K^{(2n)}=\frac 2{\beta \hbar }\Omega _S^{(2n-1)}  \label{dispari}
\end{equation}
relates the even frequency moments of $K_q$ with the odd ones of $S_q.$ The
same memory function framework of Eq. (\ref{fqz}) can be phrased for the
Kubo relaxation function and for its Laplace transform $\widetilde{K}_q(Q,z)$%
.

By virtue of all these properties, in a situation where the quantum aspects
not associated with detailed balance are marginal, it is reasonable
(although not strictly rigorous) to identify the spectrum $K_q(Q,\omega )$
with the classical quantity $S_{cl}(Q,\omega )$ so that

\begin{equation}
S_{q}(Q,\omega )\simeq \frac{\beta \hbar \omega }{1-e^{-\beta \hbar \omega }}%
S_{cl}(Q,\omega )  \label{squantclass}
\end{equation}

Having established such a correspondence, from now on we will drop out the
subscript {\it cl} and refer to the classical quantities as in fact done at
the beginning of this section.

\subsection{The memory function features\label{memoryfeatures}}

The memory function $M(Q,t)$ accounts for all the relaxation mechanisms
affecting collective dynamics, and consequently is the central quantity in
most theoretical approaches. From straightforward algebra, the initial value
of $M(Q,t)$ is related to the spectral moments of $S(Q,\omega )$ by:

\begin{equation}
M(Q,t=0)=\frac{\Omega _S^{(4)}(Q)}{\Omega _S^{(2)}(Q)}-\Omega _S^{(2)}(Q)
\label{mqt}
\end{equation}
An exact expression of $\Omega _S^{(4)}(Q)$ exists, but involves both the
derivatives of the interparticle potential and the pair distribution
function \cite{libroumberto}. It is usual to define $\omega _L^2(Q)=\Omega
_S^{(4)}(Q)/\Omega _S^{(2)}(Q)$ and $\Delta ^2(Q)=\omega _L^2(Q)-\omega
_0^2(Q)$, their meaning being evident from the following argument. For
sufficiently large $\left| z\right| $, $\widetilde{M}(Q,z)\simeq M(Q,t=0)/z$
and Eq. (\ref{cqz}) is seen to have poles at $z=\pm i\sqrt{\omega
_0^2(Q)+\Delta ^2(Q)}=\pm i\omega _L(Q)$, showing that the frequency $\omega
_L(Q)$ characterizes the instantaneous collective response of the liquid at
the wavevector $Q$. Similar remarks can be made for the generalized
infinite-frequency velocity $c_\infty (Q)\equiv \omega _L(Q)/Q.$

In liquid systems, in the long time limit, one expects that $%
M(Q,t\rightarrow \infty )$ approach zero value, therefore, regardless the
details of its shape, $M(Q,t)$ it is expected to decay over a certain
timescale $\tau (Q)$ that, for the sake of simplicity, can be defined as $%
\tau (Q)={\rm M}(Q)/\Delta ^2(Q),$ being ${\rm M}(Q)=\int_0^\infty M(Q,t)dt.$
It can be of considerable interest to point out the asymptotic behaviors of
Eq. (\ref{langevin}) in the opposite regimes $\tau (Q)\omega _0(Q)<<$ or $>>1
$. In the first limit ($M(Q,t)\approx 2{\rm M}(Q)\delta (t)$) Eq. (\ref
{langevin}) reduces to the equation for a {\it damped harmonic oscillator},
and it can be easily proved that $M^{\prime }(Q,\omega )=$ ${\rm M}(Q)$ and $%
M^{\prime \prime }(Q,\omega )=0,$ which means two inelastic peaks in the $%
S(Q,\omega )$ (in the current spectra they are centered at $\pm \omega _0(Q)$%
) damped with a factor ${\rm M}(Q).$ In the opposite limit, $\tau (Q)\omega
_0(Q)>>1,$ the decay of $F(Q,t)$ is much faster than $\tau (Q)$ and $M(Q,t)$
appearing in the convolution integral of Eq. (\ref{langevin}) can be
considered constant, $M(Q,t)\approx ${\it \ }$\Delta ^2(Q)$. Therefore Eq. (%
\ref{langevin}) becomes non homogeneous and the solution of Eq. (\ref
{sqwgenerale}) reduces to a {\it harmonic oscillator} of frequency $\omega
_L(Q)$ with no damping, plus a {\it sharp elastic line}. In this extreme
case, the intensity ratio between the elastic and inelastic lines (the Debye
Waller factor) in $S(Q,\omega )$ is $f(Q)=1-\omega _0^2(Q)/\omega _L^2(Q)$.

The true time dependence of the memory function is a priori unknown, and
from the very start it is convenient to separate in $M(Q,t)$ the decay
channels which explicitly involve couplings to thermal fluctuations ( $%
M_{th}(Q,t)$ ) from those directly associated with longitudinal density
modes ( $M_{L}(Q,t)$ ). A convenient way to perform this splitting is by the
use of the generalized hydrodynamics theory \cite{libroumberto,BY}. In the
simplest version of this approach, one writes

\begin{eqnarray}
M(Q,t) &=&M_L(Q,t)+M_{th}(Q,t)  \label{memory} \\
&=&\Delta _L^2(Q)m_L(Q,t)+\Delta _{th}^2(Q)m_{th}(Q,t)  \nonumber
\end{eqnarray}
with $\Delta _L^2(Q)\equiv \omega _L^2(Q)-\gamma (Q)\omega _0^2(Q)$, $\Delta
_{th}^2(Q)=[\gamma (Q)-1]\omega _0^2(Q)$. Here $\gamma (Q)$ is a
generalization of the specific heat ratio $\gamma =C_P/C_V$ to finite
wavevectors. The relaxation processes underlying the dynamics are accounted
for by the normalized quantities $m_L(Q,t)$ and $m_{th}(Q,t)$, defined in
such a way that $m_L(Q,0)=m_{th}(Q,0)=1$. A straightforward generalization
of ordinary hydrodynamics suggests for the thermal contribution the
following form

\begin{equation}
m_{th}(Q,t)\approx exp[-a(Q)Q^{2}t]  \label{memoryth}
\end{equation}
where $a(Q)$ can be viewed as a finite $Q$ {\it generalization} of the
quantity $D_{T}=\kappa /nC_{v},$ being $\kappa $ the thermal conductivity.
The role of the thermal contribution in the dynamic structure factor of
liquid metals will be discussed in the following section.

In contrast with the thermal decay channel, no guidance for $M_{L}(Q,t)$ is
provided by ordinary hydrodynamics. In the latter, one implicitly assumes
that

\begin{equation}
M_{L}(Q\rightarrow 0,t)\approx 2(\eta _{L}/nm)Q^{2}\delta (t)
\label{memoryL}
\end{equation}
where the longitudinal viscosity coefficient $\eta _{L}$ is related to the
ordinary shear and bulk viscosities by $\eta _{L}=(4/3)\eta +\eta _{B}$.
Clearly in the limit $Q\rightarrow 0$ the ratio $\widetilde{M}%
_{L}(Q,z=0)/Q^{2}$ approaches $\eta _{L}/nm$.

The simplest way to go beyond the hydrodynamic result (\ref{memoryL}) is to
allow for a finite decay rate of $M_L(Q,t)$:

\begin{equation}
M_L(Q,t)=\Delta _L^2(Q)e^{-t/\tau (Q)}  \label{memoryL1}
\end{equation}
where it has been additionally assumed an exponential lineshape for the
memory function. Although this has the advantage of analytical simplicity
when dealing with Fourier transform, a drawback of this ansatz lies in the
violation of some basic short-time features of the memory function (such as
a zero derivative at $t=0$), causing the divergency of $\Omega _S^{(n)}$ for
$n\geq 6.$

Eq. (\ref{memoryL1}) yields the so-called {\it viscoelastic model} for $%
S(Q,\omega )$ \cite{loves}. Since as $Q\rightarrow 0$ $\widetilde{m}%
_L(Q,z=0)/Q^2$ can written as $[c_\infty ^2-c_0^2]\tau (Q\rightarrow 0)$,
the requirement that this coincides with $\eta _L/nm$ shows that the time $%
\tau (Q)$ must be finite as $Q\rightarrow 0$. Such a connection with viscous
effects justifies the physical interpretation of the rate $1/\tau (Q)$ as a
parameter giving an overall account of all relaxation processes by which the
longitudinal response of the liquid is affected by time-dependent
disturbances. In particular, for slow perturbations developing over a
timescale $t\gg \tau (Q)$ the system can adjust itself to attain local
equilibrium and the usual viscous behavior. In contrast, for disturbances
fast enough that $t\ll \tau (Q)$ the liquid responds instantaneously, with a
solid-like (elastic) behavior. The crossover between these limiting
situations (times $t\approx \tau (Q)$, or frequencies $\omega $ such that $%
\omega \tau (Q)\approx 1$) is ultimately responsible for the gradual changes
often detected in the sound dispersion of several liquids at increasing
wavevectors.

Similar considerations can be applied to the $M_{th}(Q,t)$ contribution
which involves the timescale $\tau _{th}=D_T^{-1}(Q)Q^2;$ in this case the
''elastic response '' is achieved in the hydrodynamic regime: $\omega
_0(Q\rightarrow 0)\tau (Q\rightarrow 0)\propto c_0/Q\rightarrow $ $\infty ,$
while the ''viscous response '' emerges at short wavelengths. The crossover
between the two limits embodies the transition between the adiabatic and
isothermal regimes, and reflects the actual possibility of density
fluctuations to decay by the thermal channel: at high frequencies there is
no time for such conversion, and the fluctuations evolve without heat
transfer (higher sound speed, no attenuation) while in the low frequency
limit there is enough time for the system to equilibrate so that an
isothermal dynamics takes place (lower sound speed, damped excitation).

These qualitative considerations are supported by the evolution of the poles
of $\widetilde{C}_{L}(Q,z)$ at increasing wavevectors. Eqs. (\ref{memory}), (%
\ref{memoryth}) and (\ref{memoryL1}) in fact imply that

\begin{equation}
\widetilde{M}(Q,z)=\frac{\Delta _L^2(Q)}{z+1/\tau (Q)}+\frac{\Delta
_{th}^2(Q)}{z+a(Q)Q^2}  \label{m2tempi}
\end{equation}

In the hydrodynamic regime ($Q\rightarrow 0$) both the inequalities $\left|
z\right| \ll 1/\tau (0)$ and $\left| z\right| \gg $ $a(Q)Q^2$ are satisfied,
and the poles are approximately located at $z\approx \pm i\sqrt{\omega
_0^2(Q\rightarrow 0)+\Delta _{th}^2(Q\rightarrow 0)}=\pm i$ $c_s$ $Q$, where
$c_s\equiv \sqrt{\gamma }$ $c_0$ is the adiabatic sound velocity. At larger
wavevectors different situations may arise: the one of our interest is $%
\left| z\right| \ll a(Q)Q^2$ , which is readily seen to cause a shift of the
imaginary part of the poles toward an isothermal behavior, i.e. $\omega
\simeq \pm \omega _0(Q)$. The last limiting case occurs whenever the
frequency $\omega $ becomes distinctly larger than $\left[ 1/\tau (Q)\right]
$, which eventually yields poles at $z=\pm \omega _L(Q)$, as already
noticed. Although appealing, the simplicity of the viscoelastic model can be
deceptive. First of all, the model itself provides no clue for the physical
origin of the decay mechanisms leading to the rate $1/\tau (Q)$. Secondly,
even ignoring this aspect and treating $1/\tau (Q)$ as a fitting parameter,
the practical results of the model are rather unsatisfactory (see Fig. \ref
{3comparison} in the following). These drawbacks of the simple viscoelastic
model have been theoretically predicted, and somehow reported in a number of
liquids through a detailed analysis of MD spectra \cite{libroumberto,casasMC}%
. However, to our knowledge, no cutting edge analysis exists in the case of\
experimental measurements of the coherent dynamic structure factor, probably
because of the afore-mentioned limits of the neutron technique.

The obvious remedy is to modify the simple ansatz (\ref{memoryL1}) by
allowing a more sophisticated decay of $M_L(Q,t)$. We adopted the following
two-exponential ansatz:

\begin{equation}
M_{L}(Q,t)=\Delta _{L}^{2}(Q){\ }\left[ {(1-\alpha (Q))e}^{{-\gamma }_{1}{%
(Q)t}}{+\alpha (Q)e}^{{-\gamma }_{2}{(Q)t}}\right]  \label{x2tempi}
\end{equation}
where the rate ${\gamma }_{1}{(Q)}$ is chosen to be larger than ${\gamma }%
_{2}{(Q)}$, so that the dimensionless factor ${\alpha (Q)}$ measures the
relative weight of the ''slow'' decay channel. Besides being more flexible
than the viscoelastic model, we shall see that the ansatz (\ref{x2tempi})
has the much more important merit that the presence of two different
timescale does in fact have a definite physical interpretation.

The previous ansatz has been proposed phenomenologically several years ago
and tested against the MD results in LJ fluids \cite{leves}, it is
interesting to note that an expression analogous to Eq. (\ref{x2tempi}) has
been implicitly introduced in the viscoelastic analysis of Brillouin Light
Scattering (BLS) spectra of glass forming materials (see for example ref.
\cite{cumm}). In fact, in these works, the general expression of $M_L(Q,t)$
for a two times exponential decay is always expressed as

\begin{equation}
M_L(Q,t)=\Delta _\mu ^2(Q){e}^{{-t/\tau }_\mu }+\Delta _\alpha ^2(Q){e}^{{%
-t/\tau }_\alpha }  \label{BLS}
\end{equation}
with explicit reference to the so-called $\alpha $- (structural) relaxation
process as responsible for the long lasting tail, and to the $\mu -$
microscopic process as additional, faster, relaxation dominant over a very
short timescale. To be more precise, in the BLS spectral window the
condition $\omega \tau _\mu <<1$ holds, so that the approximation

\[
M_L(Q,t)={2}\Delta _\mu ^2(Q)\tau _\mu {\delta (t)+}\Delta _\alpha ^2(Q){e}^{%
{-t/\tau }_\alpha }
\]
is customarily adopted as a fitting model. On the contrary, we will show in
this work how such an approximation is no longer tenable in the case of
liquid Lithium at the IXS frequencies. The origin, at the atomic level, of
this fast decay channel is still an open issue: one of the purposes of the
present paper is to point out the features of such relaxation and to inspect
its nature. The rapidly decaying portion of $M_L(Q,t)$ is customarily
attributed to largely uncorrelated collisional events, similar to those
occurring in a dilute fluid. In addition, at the high densities typical of
the liquid state, non-negligible correlations among the collisions can be
expected, making no longer valid an interpretation only in terms of
''binary'' collisions. Albeit the magnitude of the correlation effects is
relatively small and their buildup slow, once established their decay is
even slower, and for $t>1/{\gamma }_1{(Q)\equiv \tau }_\mu $ this relaxation
channel dominates the decay of $M_L(Q,t)$, which consequently may exhibit a
small but long-lasting ''tail''. The ansatz (\ref{x2tempi}) can incorporate
most of this physics: on the basis of the latter, one may reasonably
anticipate that ${\alpha (Q)\ll }1$, and that the time $1/{\gamma }_2{%
(Q)\equiv \tau }_\alpha $ is distinctly longer than $1/{\gamma }_1{(Q)\equiv
\tau }_\mu $. In this picture, the best fitted values of the viscoelastic
rate $1/\tau (Q)$ clearly represent some sort of ''weighted average''
between ${\gamma }_1{(Q)}$ and ${\gamma }_2{(Q)}$. Finally, we may argue
that at increasing $Q$ (namely, over a shrinking length scale) the magnitude
${\alpha (Q)}$ of correlation effects should decrease, and that at higher
temperatures the value of ${\alpha (Q)}$ at a given wavevector should
equally decrease. On a general basis, the requirement that $%
\lim_{Q\rightarrow 0}M_L(Q,z=0)/Q^2\rightarrow \eta _L/nm$ now takes the form

\begin{equation}
(c_\infty ^2-c_0^2)\left[ \frac{{(1-\alpha (Q\rightarrow 0))}}{{{\gamma }_1{%
(Q\rightarrow 0)}}}{\ +\ }\frac{{\alpha (Q\rightarrow 0)}}{{{\gamma }_2{%
(Q\rightarrow 0}\ )}}\right] \rightarrow \eta _L/nm  \label{limite}
\end{equation}
The refined model (\ref{x2tempi}) yields a dynamic structure factor given by

\begin{eqnarray}
&&S(Q,\omega )=\frac{S(Q)}\pi  \label{x3tempi} \\
&&\ \times Re\left\{ \frac{\omega _0^2(Q)}{i\omega +\frac{\Delta _\mu ^2(Q)}{%
i\omega +{\gamma }_1{(Q)}}+\frac{\Delta _\alpha ^2(Q)}{i\omega +{\gamma }_2{%
(Q)}}+\frac{\Delta _{th}^2(Q)}{i\omega +{a(Q)Q}^2}}\right\} ^{-1}.  \nonumber
\end{eqnarray}

\section{THE\ EXPERIMENTS}

In this work we report the determination by IXS of the dynamic structure
factor of liquid lithium in the wavevector range from $1.4$ to $110$ nm$%
^{-1} $ , corresponding to $Q/Q_{m}\approx 5\cdot 10^{-2}\div 5.$ The
scanned energy range has been settled at each $Q$ in order to detect all the
scattered signal up to the tails region, where it becomes comparable to the
background.

The experiment has been performed at the very high energy resolution
beamline ID16-BL21 at the European Synchrotron Radiation Facility. A
monochromatic beam of $10^{9\text{ }}$photons/s is obtained from a
cryogenically cooled Si(111) double crystal followed by a high energy
resolution monochromator operating in back-scattering geometry at selectable
Bragg reflections \cite{noiV}. The scattered beam is collected by perfect
spherical silicon crystal analyzers operating in back-scattering and Rowland
circle geometry at the same reflection order of the monochromator. They were
obtained by gluing $\simeq 12000$ perfect crystals of $0.6\times 0.6\times 2$
mm$^3$ on a spherical blank \cite{noiM}. The overall energy resolution has
been measured using a plexiglass scatterer at a $Q$\ value corresponding to
the maximum of its structure factor where the diffusion is dominated by the
elastic contribution, and it turns out to have approximately a Lorenzian
shape.

Different configurations have been adopted depending on the explored $Q$%
-range. Below $Q_m$ ($25$ nm$^{-1}$) we used simultaneously five-analyzers
mounted on a horizontally rotating arm and with a fixed offset of $1.5^0$.
Either the Si(9 9 9) or Si(11 11 11) reflections, corresponding to energy
resolutions of $3.0$ meV and $1.5$ meV (FWHM) respectively, have been
selected. At higher $Q$ values, where the spectra start to broaden, we used
a vertical scattering geometry based on a single analyzer housed on a
rotating arm that can reach more than 150$^0$ scattering angle. In this
configuration we selected the Si (7 7 7) reflection with an overall energy
resolution of $8.5$ meV FWHM. The $Q$ resolution was $\simeq \pm 0.2$ nm$%
^{-1}$ and $\simeq \pm 0.35$ nm$^{-1}$ in the horizontal and vertical
geometries respectively. The momentum transfer is related to the scattering
angle by the relation $Q=2k_0sin(\theta _s/2)$.

Energy scans have been done by varying the temperature of the monochromator
with respect to that of the analyzer crystals. The absolute energy
calibration between successive scans is better than $1$ meV. Each scan took
about 3 hours, and each $Q$-point spectrum has been obtained from the
average of 2 to 8 scans depending on the Bragg-reflection order and on the $%
Q $-transfer. The data have been normalized to the intensity of the incident
beam. In the $Q-E$ region of interest, empty vacuum chamber measurements
gave either the flat electronic detector background of 0.6~counts/min or, in
the angular region corresponding to $9<Q<13$ nm$^{-1}$, a small elastic line
due to spurious reflection of the beam entering the analyzers. These background
signals have been subtracted from the data before any other data treatement.
The chamber kapton windows (each 50 $\mu $m thick) gave no detectable contribution
to the scattered flux.

The liquid lithium uncapped container was made out of austenitic stainless
steel in contact with a resistive heater, used to kept the liquid at
constant temperature. We selected two different values: one slightly above
the melting point, $T=475$ K, and the second at $T=600$ K; we have not been
able to reach higher temperatures with our experimental setup due to the
lack of windows. The $20$ mm long sample, kept together by surface tension,
was maintained in a 10$^{-6}$ bar vacuum. The lithium has been loaded in an
argon glove box.

As far as experimental aspects are concerned, the {\it energy transfer} $E$
between the probe (X-rays) and the sample (excitations) has always been
measured in meV. To make contact with the traditional notations adopted in
computer simulations we will refer from now on only to the frequency $\omega
=2\pi \upsilon =E/\hbar $ expressed in ps$^{-1}$ (a numerical factor of $%
\approx 0.659$ ps$^{-1}$/meV accounts for such a conversion).

\begin{figure}[h]
\centering
\includegraphics[width=.8\textwidth]{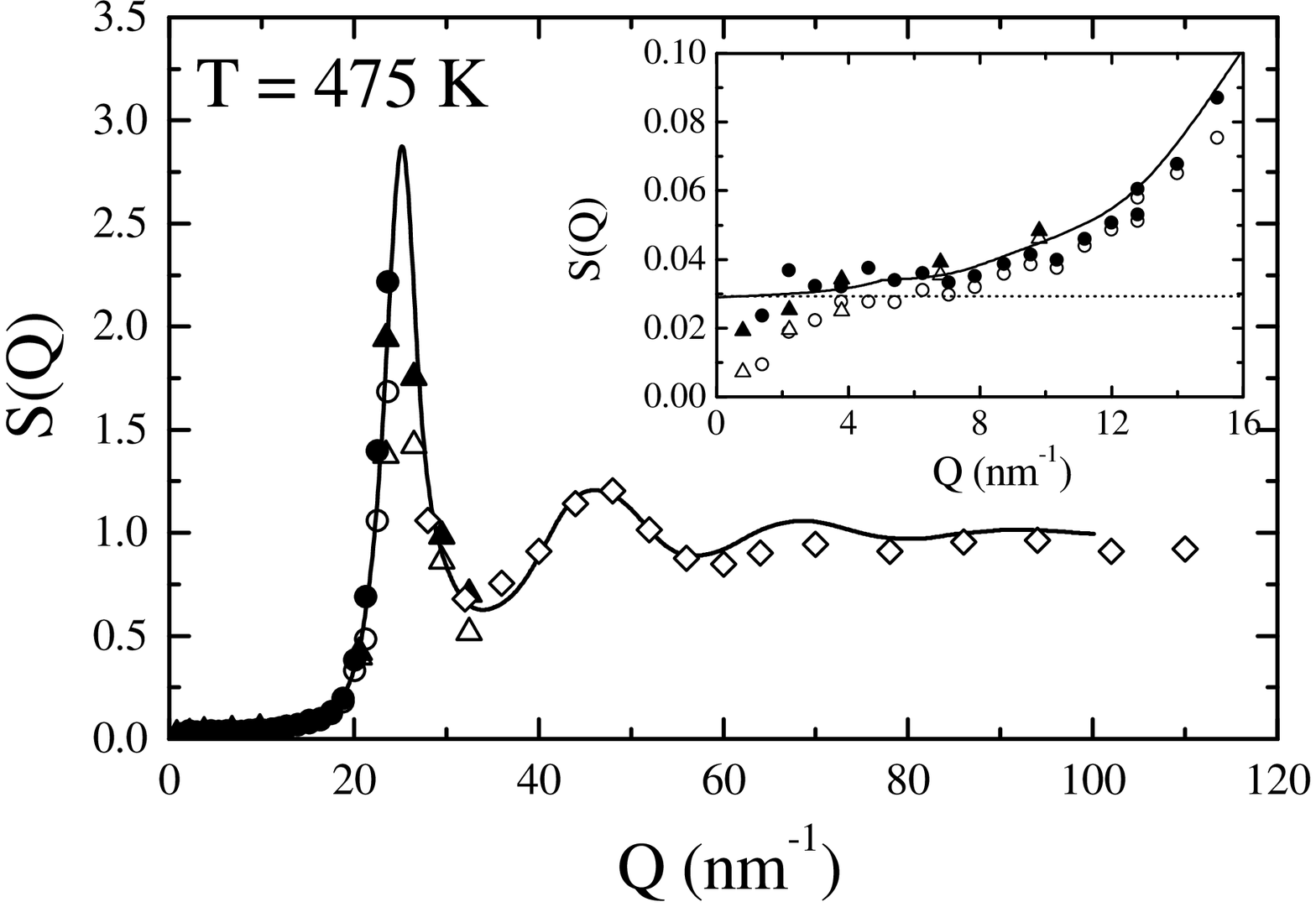}
\vspace{-9cm}
\caption{Comparison between the $S(Q)$ experimentally determined as
discussed in the text and trough MD simulations (------) of ref.[28].
Triangle ($\triangle $): (11 11 11) data, Dots: ($\circ $) (9 9 9) data,
Lozenges ($\diamondsuit $): (7 7 7) data. The full symbols are the values
corrected for asymmetry resolution effects, at the higher Q's such
correction is negligible and is not reported. The inset shows on an enlarged
scale the low $Q$ region.}
\label{sq}
\end{figure}

The IXS spectra of liquid lithium below $Q_m$ (horizontal geometry) are
reported in Fig. \ref{IXSlowq} at the indicated $Q$ values. The low $Q$ data
show the Brillouin triplet structure with the energy of the inelastic peaks
increasing with $Q$ up to a $Q$ value of $12$ nm$^{-1}$. This value is close
to $Q_m/2$, as deduced from the $S(Q)$ reported in Fig. \ref{sq}. The
dispersion up to $Q_m/2$ can therefore be interpreted as that of the
longitudinal acoustic branch in a pseudo first Brillouin zone (BZ).
Furthermore, similarly to what is found in the second BZ of a crystal, we
observe that, also in liquid lithium, the energy of the acoustic modes
decreases with increasing $Q$ from $Q_m/2$ to $Q_m$. Increasing $Q$ above $%
Q_m$, i.~e. in the ''third'' or higher BZs, the spectrum increasingly
broaden and distinct peaks are no longer observable. At the highest $Q$%
-values one finds that the dynamic structure factor becomes a symmetric peak
centered at frequencies larger than $\omega =0,$ the behavior expected for a
quantum free particle (in the classical limit no recoil energy is expected).
Such evolution can be clearly followed in Fig. \ref{IXShighq}. Beside the
observation of dispersion in a first and second pseudo-BZs, it is also
important to note that the broadening of the excitations increases
monotonically with $Q$,\ to the extent that near the end of the second
pseudo-BZ a well defined inelastic peak is no longer visible.

In order to extract quantitative information from the experimental
intensity, i.e. to perform measurements of $S_q(Q,\omega )$ on an absolute
scale, the most direct way would be to use a reference scatterer as
customarily done in neutron experiments. In IXS such a procedure is
extremely difficult because of the $Q$-dependence of the form factor and, in
our case, for the different efficiencies of the analyzers. For these reasons
we preferred to use an alternative indirect method based on the knowledge of
the sum rules of $S_q(Q,\omega )$: in particular for the first two frequency
moments we have
\begin{eqnarray*}
\Omega _S^{(0)} &=&\int S_q(Q,\omega )d\omega =S(Q), \\
\Omega _S^{(1)} &=&\int \omega S_q(Q,\omega )d\omega =\hbar Q^2/2m.
\end{eqnarray*}
where the second equality follows from Eq. (\ref{dispari}) applied for $n=1$%
. The measured raw intensity is related to the dynamic structure factor
through

\begin{equation}
I(Q,\omega )=A(Q)\int d\omega ^{\prime }S_q(Q,\omega ^{\prime })R(\omega
-\omega ^{\prime })  \label{convo}
\end{equation}
where $R(\omega )$ is the experimental resolution function and $A(Q)$ is a
factor taking into account the scattering geometries, the experimental setup
and the lithium atomic form factor. The first moments of the experimental
data, $\Omega _I^{(0)}$ and $\Omega _I^{(1)}$, and those of the resolution
function, $\Omega _R^{(0)}$ and $\Omega _R^{(1)}$, are related to $\Omega
_S^{(0)}$ and $\Omega _S^{(1)}$ by:

\begin{eqnarray*}
\Omega _{I}^{(0)} &=&A(Q)\Omega _{S}^{(0)}\Omega _{R}^{(0)}, \\
\Omega _{I}^{(1)} &=&A(Q)(\Omega _{S}^{(0)}\Omega _{R}^{(1)}+\Omega
_{S}^{(1)}\Omega _{R}^{(0)}).
\end{eqnarray*}
From the previous equation one derives that
\begin{equation}
S_{q}(Q)=\frac{\hbar Q^{2}}{2M}(\Omega _{I}^{(1)}/\Omega _{I}^{(0)}-\Omega
_{R}^{(1)}/\Omega _{R}^{(0)})^{-1}.  \label{norma}
\end{equation}

This procedure has been adopted to establish an absolute scale for $%
S_q(Q,\omega )$ using the experimentally determined $I(Q,\omega )$ and $%
R(\omega )$. Its reliability is shown in Fig. \ref{sq} where we obtain an
excellent agreement between the $S_q(Q)$ values obtained by Eq. (\ref{norma}%
) and those derived by a MD simulation at $T=475$ K \cite{tullioMD,notasq}.
The reported MD data are the outcame of a simulation made using the Price-Tosi
interaction pseudopotential already tested against neutron diffraction data
(see ref \cite{umblitiomd} for further details).

\section{DATA\ ANALYSIS\label{analisi}}

Lithium has been chosen because, among the simple monatomic liquids, is the
one that is best suited to IXS. Indeed, its low mass gives recoil energies
observable in the considered $Q$-range, and its low atomic number and large
sound velocity give optimal signal within the available energy resolution,
compensating for the large form factor decrease at high $Q$ values. The $%
I(Q,\omega )$ spectra, reported to their absolute scale exploiting the
zeroth and first moments sum rules, show the transition from a triplet to a
Gaussian. The maxima, $\omega _{m}(Q)$, of the longitudinal current spectra (%
$\omega ^{2}S(Q,\omega )/Q^{2}$) show an almost linear dispersion relation
at low $Q$ - typical of a sound wave - and a completely different dependence
in the high $Q$ limit, where they approach the parabolic dispersion of the
quantum free particle. Between these two regions, $\omega _{m}(Q)$ exhibits
oscillations which are in phase with the structural correlations as observed
in $S(Q)$.

The low $Q$ region is of particular interest for its large number of
informations. The so called positive dispersion, a fingerprint of the
relaxation dynamics of disordered systems, either liquid \cite
{umblitiomd,canaleslitiomd,rubidio,rubidio2,cesio} or glassy \cite
{silicamasciove,tesitullio}, reflects the host of mechanisms driving such
dynamics, their timescale and, at a careful inspection, their nature. We
will start our analysis from here.

\subsection{The ''low Q region''}

\begin{figure}[h]
\centering
\includegraphics[width=.8\textwidth]{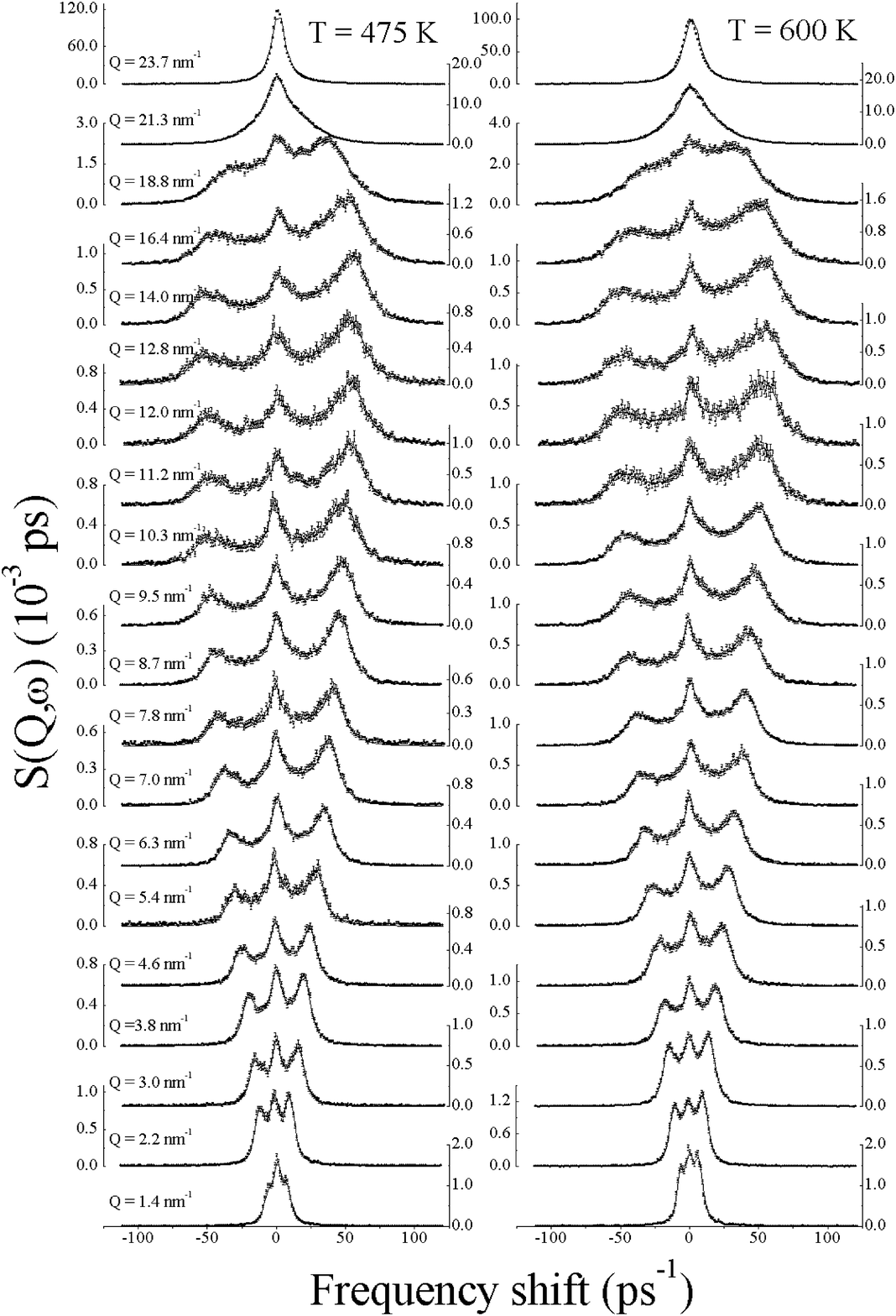}
\caption{IXS spectra at selected $Q<Q_{m}$ values, in the (9 9 9)
configuration, at two different temperatures. The full line is the best fit
taking into account thermal, slow and fast relaxations.}
\label{IXSlowq}
\end{figure}

\begin{figure}[h]
\centering
\includegraphics[width=.8\textwidth]{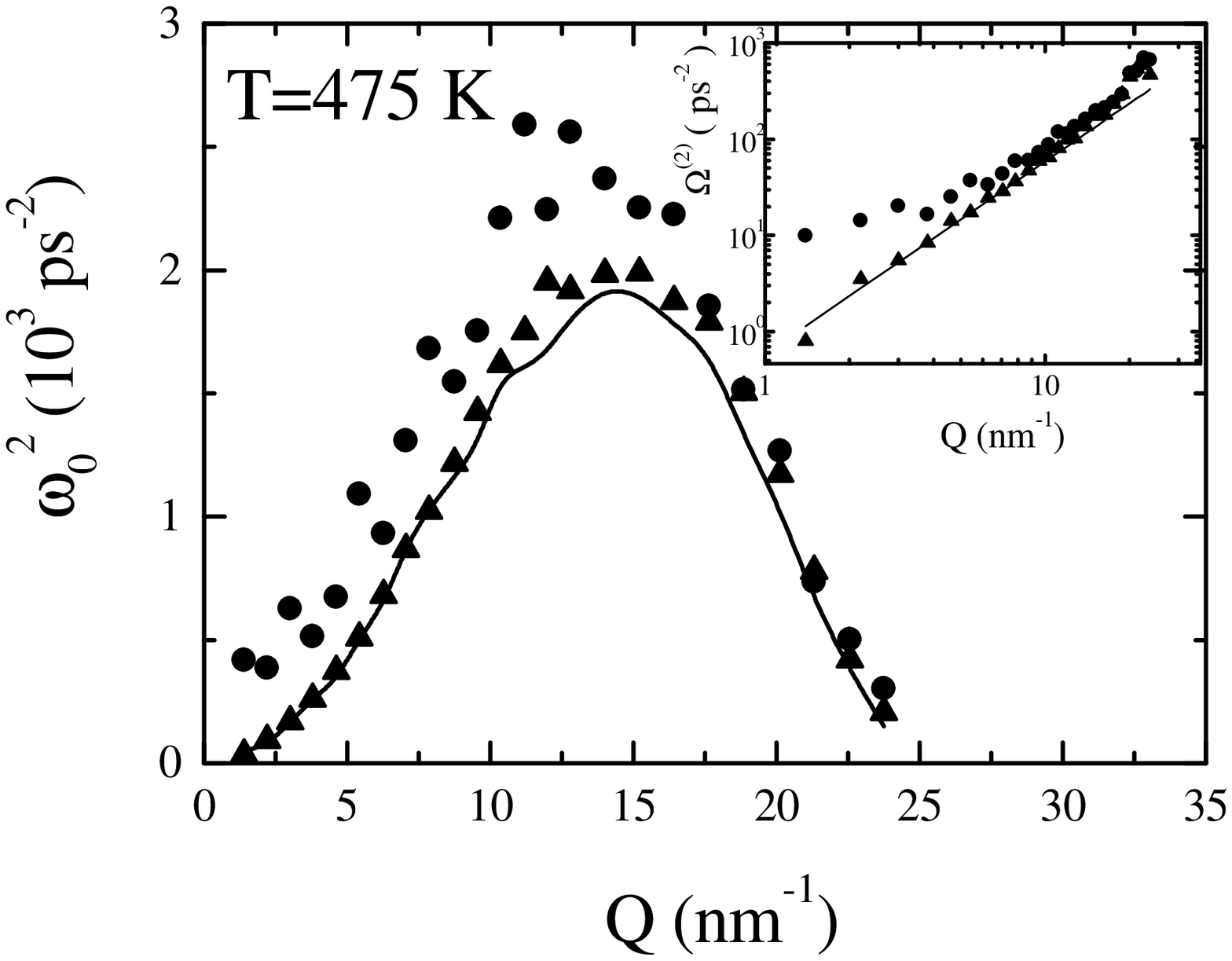}
\vspace{-9cm}
\caption{Comparison between the value of $\omega _0^2(Q)$ as calculated from
its definition (------) and from the three times fit ($\blacktriangle $)
leaving it as a free parameter. The normalized second frequency moment of
I(Q,$\omega $) is also reported ($\bullet $) to emphasize resolution and
quantum effects. In the inset we show the same non-normalized quantities.}
\label{msecondo}
\end{figure}

Looking at the spectra in Figs. \ref{IXSlowq},\ref{IXShighq}, as also
confirmed by the results reported in this section, the memory function
approach can profitably be used approximately up to $Q_m.$ We adopted the
following procedure. The classical result of Eq. (\ref{sqwgenerale}) has
been modified according to Eq. (\ref{squantclass}) to take into account of
the detailed balance effects. The resulting $S_q(Q,\omega )$ has then been
convoluted with the instrumental resolution according to (\ref{convo}). The $%
\chi ^2$ function arising from the difference between the outcoming function
and the experimental data has finally been minimized by a standard
Levenberg-Marquardt routine. Among the fitting parameters, $S(Q)$ and $%
\omega _0^2(Q)$ are independent of the specific model for the memory
function, and their value is basically known: the first has been calculated
from the first moment sum rule as discussed in Sec. III (we have already
discussed the approximation $S_q(Q)\cong S(Q)$), so that it has been kept
fixed during the iterations. Then the second quantity $\omega _0^2(Q)$ is
simply deduced from its definition (see Eq. (\ref{a2})). It is worth to
point out that $\omega _0^2(Q)$ {\it cannot be taken as the second frequency
moment of }$I(Q,\omega )$ for two main reasons. The first one, related to
the effects of the resolution, may be in principle overcome by a procedure
similar to the one leading to the determination of $S(Q),$ but with the
further difficulties arising from the almost-Lorenzian shape of the
resolution that leads to a diverging $\Omega _R^{(2)}$. Most severe is the
{\it non-invariance of the second moment under the transformation }(\ref
{squantclass}). For the sake of clarity we report in Fig. \ref{msecondo} the
theoretical value $\omega _0^2(Q)$ compared to the second normalized moment
of the experimental $I(Q,\omega )$ at $T=475$ K: the latter is always larger
(this may be due also to the small but non-vanishing background in the
experimental data) and the agreement gets worse at decreasing wavevectors,
when the resolution effects become dominant.

All the other fitting parameters obviously depend on the specific model for $%
M(Q,t)$. Let us firstly examine the spectral relevance in liquid Li of the
coupling to thermal fluctuations, accounted for by the contribution $%
M_{th}(Q,t)$ to the total memory function. At not too high temperatures in
the molten phase, all liquid alkali metals are characterized by a specific
heat ratio $\gamma \approx 1,$ the typical values giving $\gamma -1\simeq 0.1
$. Outside the hydrodynamic ($Q\rightarrow 0$) regime (and, in particular,
in the wavevector range probed in this work) there are several indications
that the difference $\gamma (Q)-1$ is even smaller \cite
{canaleslitiomd,notacanales}. Despite its modest strength in the present $Q$%
-range, the role of thermal relaxation in liquid metals is strikingly
different from the one played in ordinary non-conducting liquids. A simple
estimate of the decay rate embodied in the $M_{th}(Q,t)$ reported in Eq. (%
\ref{memoryth}) based on the experimental values of the thermal conductivity
and of the specific heat \cite{ose} shows that, in liquid lithium, $\omega
\ll (\kappa /nC_V)Q^2$ for most of the spectral range of interest here, and
in particular for the range characterized by the presence of inelastic peaks
(see Fig. \ref{IXSlowq}). This indicates that in liquid Li (but a similar
argument is expected to hold in general for molten metals) the
afore-mentioned transition from the adiabatic to the isothermal response
occurs at wavevectors well {\it below} those probed in the present
experiments. Indeed, for wavevector-independent thermal parameters (a
reasonable assumption at so low $Q$) we find for the threshold wavevector $%
\omega _0(Q^{*})\tau _{th}(Q^{*})=1$ a value of $Q^{*}\simeq 0.2$ nm$^{-1}$
at $T=475$ K, varying only of few percents at $T=600$ K. Of course, this
does not mean that the presence of thermal fluctuations is not influent: as
shown for example by Eq. (\ref{m2tempi}), and as can be inferred by the
simple arguments of Sec. \ref{def}, they may still give some contribution to
the real part of the memory function, i.e. ultimately to Brillouin peaks
widths. Such linewidth contribution gives $\delta \omega _{th}=\Delta
_{th}^2\tau _{th}=\left( \gamma (Q)-1\right) c_0(Q)/D_T(Q)\cong 0.056$ ps$%
^{-1}$ ($0.07$ ps$^{-1}$) for $T=475$ K ($T=600$ K) (in the $Q\rightarrow 0$
limit) and is not explicitly dependent on $Q$. As already remarked, such
contribution may be of some relevance only at the first two or three $Q$
values investigated here, but {\it the strongly }$Q${\it -dependent
broadening of the inelastic components has to be ascribed to some other
dynamical process}. On the basis of the same argument, it is worth pointing
out that {\it the hydrodynamic value of the sound speed is not yet reached
at the lower Q values of this work}: as far as the positive dispersion is
concerned, in the present case, {\it the lower edge of the transition is the
isothermal rather than the adiabatic value}. As previously noted, a
completely different scenario characterizes instead non-metallic fluids: due
to the much lower thermal diffusivity, the opposite thermal regime is
experienced all over the collective excitation region and the final low $Q$
result is an adiabatic response without any damping of the sound waves and
with a narrow elastic line.

\begin{figure}[h]
\centering
\includegraphics[width=.8\textwidth]{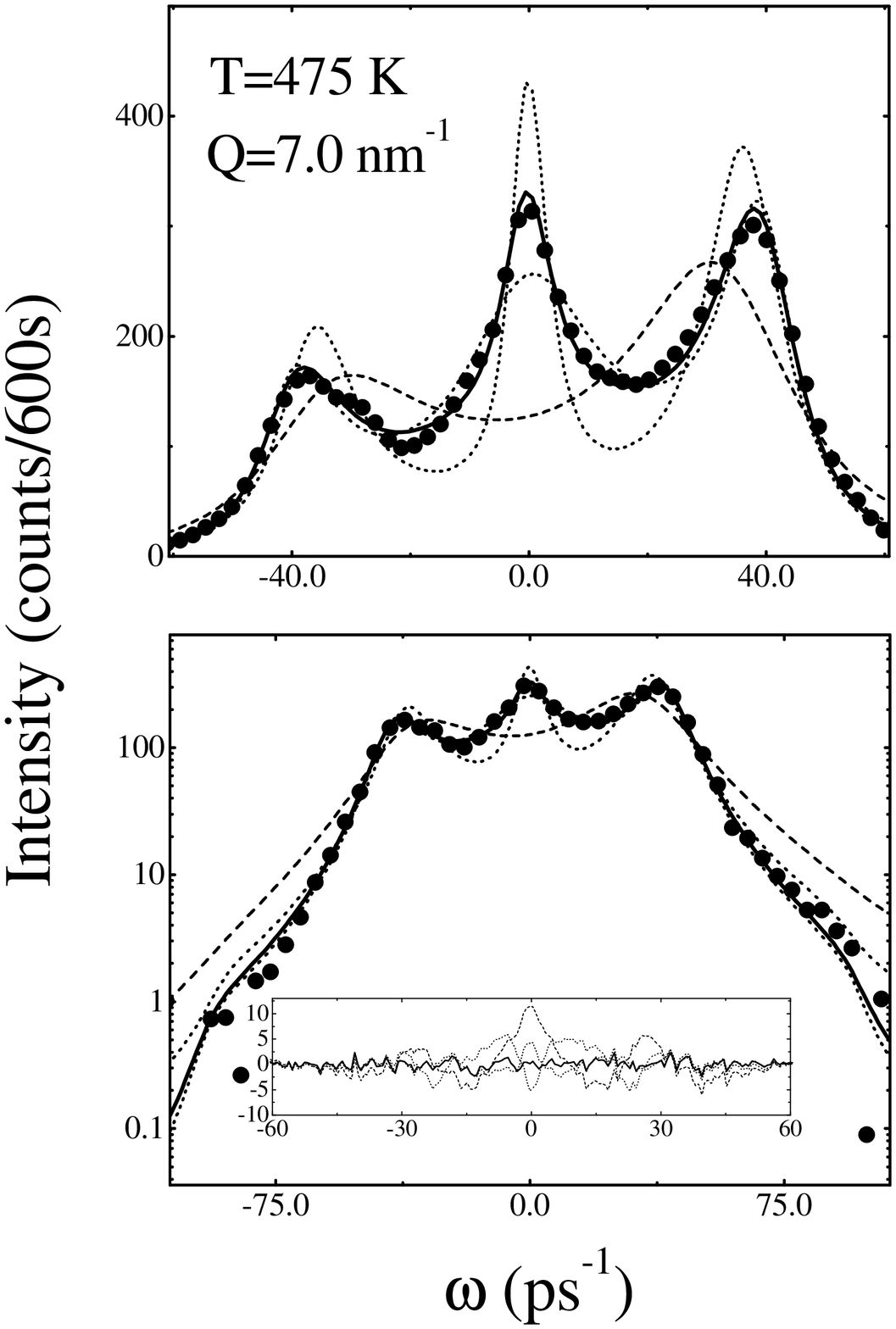}
\caption{Inelastic x-ray scattering spectrum of Lithium at $T=475$ K and $%
Q=7.1$ nm$^{-1}$ Dots ($\bullet $): experimental data; Dashed line ($---$):
simple hydrodynamic fit (DHO); Dotted lines ($\cdots \cdots $): simple
viscoelastic model (thermal and one relaxation time) showing two different
sessions of comparable $\chi ^2$; Full line (------): thermal and two
relaxation times, structural and microscopic. The inset shows the mean
square deviation of the different models.}
\label{3comparison}
\end{figure}

Having ascertained the role of thermal effects, we now come to the more
important implications of the fitting analysis for the dynamics of $%
m_{L}(Q,t)$. In the following we will refer to Fig. \ref{3comparison} where
the specific situation of the spectrum at $Q=7$ nm$^{-1}$ has been chosen to
enlighten the lineshape details associated with the different contributions
to the memory function.

As a zero-level approximation we consider the lineshape obtained by the
simple hydrodynamic expression which only accounts for thermal decaying
processes, embodying all the other relaxation channels as instantaneous:
beside the use of Eqs. (\ref{memory},\ref{memoryth}), this is simply done
assuming that $m_L(Q,t)\varpropto \delta (t)$. As in our case even the
thermal relaxation is quite rapid on the X-rays timescale, the net result is
a lineshape close to the one of the damped harmonic oscillator model. It is
therefore not surprising (see Fig. \ref{3comparison}) that the agreement
with experimental data is very poor (no elastic peak is reproduced).
Consequently, the assumption of an instantaneous decay of $m_L(Q,t)$ is
untenable.

Following section \ref{memoryfeatures}, as a next step, we generalize the
memory function by the expression (\ref{m2tempi}) i.e. the viscoelastic
model. All the merits and the drawbacks of this model are again illustrated
in Fig. \ref{3comparison}. Leaving the time $\tau (Q)$ and the strength $%
\Delta _L^2(Q)$ as fitting parameters, it is possible to obtain a rather
good agreement for the position of the inelastic peaks, and consequently for
the main features of the dispersion of sound-like excitations. However, one
notes clear discrepancies in the quasi- elastic region at small $\omega $.
The fitting procedure yields with comparable $\chi ^2$ values two different
sets of parameters, corresponding to two different timescale driving the
dynamics of the system. More precisely, in one case the broader part of the
quasielastic and the Brillouin lineshape are well reproduced but the narrow
quasi-elastic is missing, while in the other one the narrow elastic peak is
well fitted but the Brillouin lines are too sharp. In other words, {\it the
viscoelastic model cannot account for the double-sloped shape of the
quasielastic region}. Although in Fig. \ref{3comparison} we refer to $Q=7$ nm%
$^{-1},$ the situation at most wavevectors is similar.

The inescapable conclusion is that there are at least two different
timescale, each giving a contribution to the width of the ''quasi-elastic''
region, and accounting at the same time for the inelastic lineshape. On
these grounds, the more flexible phenomenological model (\ref{x2tempi}),
based on two exponential decays, appears more promising. In this case the
best fitted lineshapes (quantum corrected and resolution convoluted),
including the thermal relaxation (using the $Q\rightarrow 0$ values
of specific heat ratio and thermal diffusivity from
\cite{ose}) through Eq. (\ref{x3tempi}), are again reported in Figs. \ref
{IXSlowq} and \ref{3comparison}, and a marked improvement of the agreement
is clearly visible.

\begin{figure}[h]
\centering
\includegraphics[width=.8\textwidth]{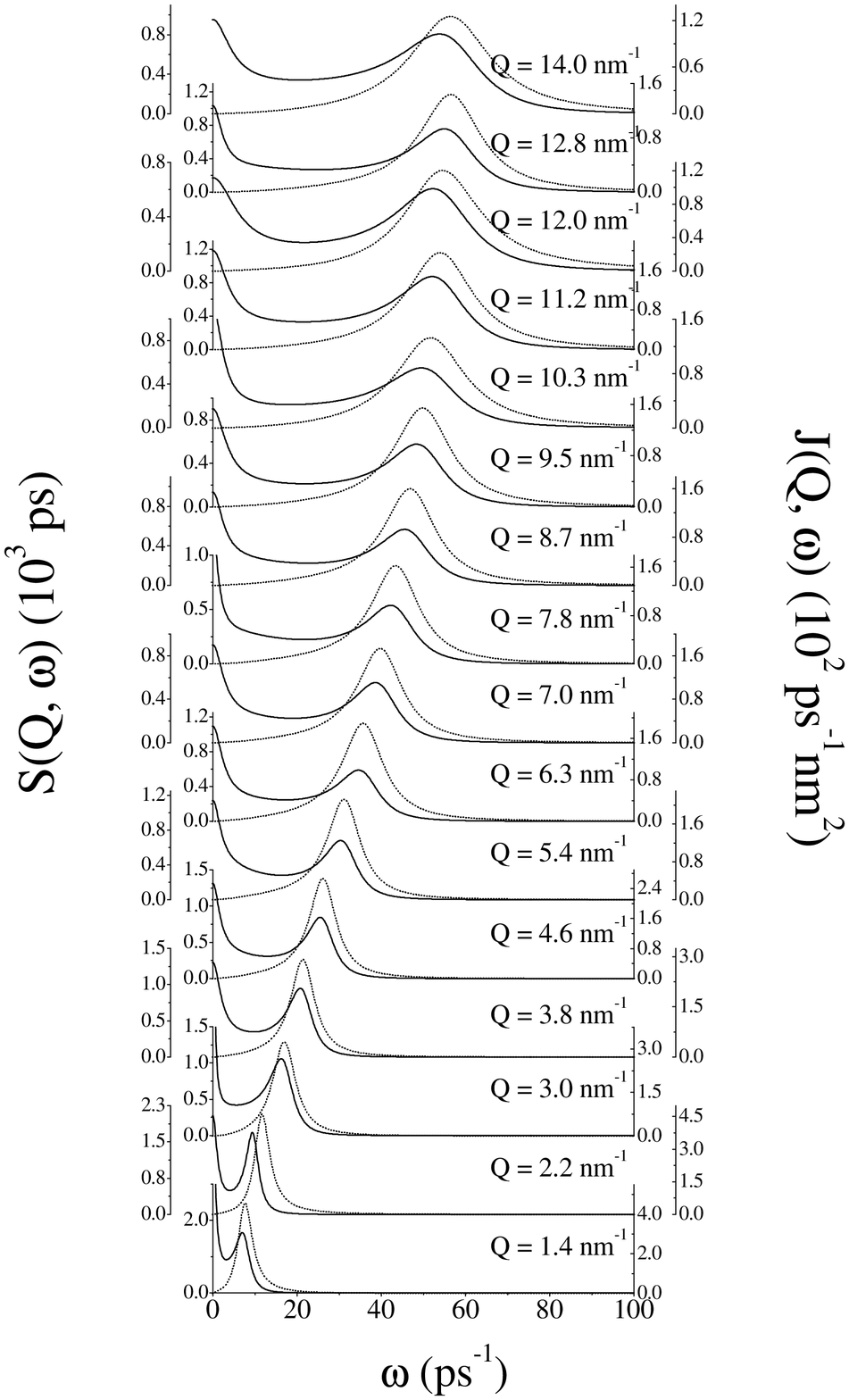}
\caption{Dynamical structure factor $S(q,\omega)$ (------) and longitudinal
current correlation function $C_L(Q,\omega)$ ($\cdots \cdots$) in the low-Q
region as obtained from the two times fitting model (classical and resolution
deconvoluted). The (symmetical) peak positions of the latter quantity have
been utilized to determine the apparent sound speed of the system.}
\label{deco}
\end{figure}

In Fig. \ref{deco} we also report the classical and resolution deconvoluted
lineshapes of $S(Q,\omega)$ and $C_L(Q,\omega)$ as builded from the fitting
parameters, in the region $Q<14$ nm$^{-1}$, the most significant as far as
the positive dispersion task is concerned. The maxima of the latter function have
been utilized to determine the apparent sound velocity of the system.

\begin{figure}[h]
\centering
\includegraphics[width=.8\textwidth]{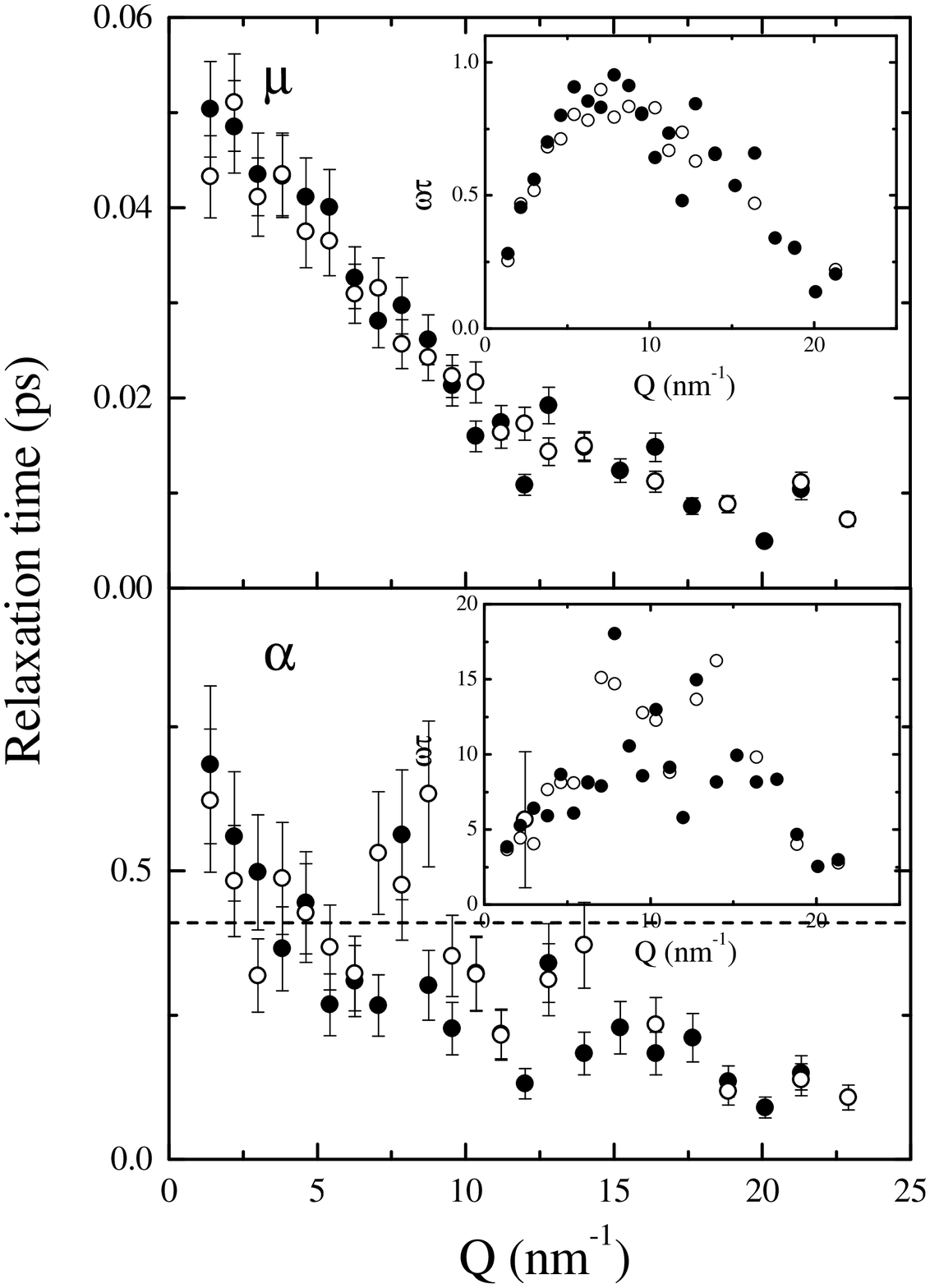}
\caption{Relaxation times at different temperatures from the fits. Full
symbol: $T=475$ K, open symbol $T=600$ K. The timescale corresponding to the
experimental resolution ($3.0$ meV) is also reported ($---$) to show how it
could affect the lower relaxation time determination.}
\label{figtimes}
\end{figure}

Let us now discuss the different parameters entering Eq. (\ref{x2tempi}),
starting from the relaxation times: in Fig. \ref{figtimes} we report the
wavevector dependence of such quantities at both temperatures. The presence
of the double timescale is confirmed by the difference by about one order of
magnitude between the two parameters $\tau _\alpha $ and $\tau _\mu .$
Concerning $\tau _\alpha $ it is worth to point out the non negligible
effect of resolution: the value of $1/HWHM$ is comparable to the relaxation
time itself so that it is expected to affect this fitting parameter,
particularly at low $Q$. As a consequence it is unfortunately not possible
to inspect any temperature dependence of $\tau _\alpha (Q,T).$ Completely
different is the situation as far as the fast timescale is concerned. The
determination of $\tau _\mu $ is expected to be definitely more reliable and
on the basis of our analysis the fast dynamics turns out to be temperature
independent at the investigated $Q$-values. The influence of these
relaxation mechanisms on the dynamics can be inferred looking at the $\omega
(Q)\tau (Q)$ values as done before for the thermal channel: as far as the
slow mechanism is concerned the system response turns out to be ''solid
like'' across the full first Brillouin pseudo-zone (see the insets in Fig.
\ref{figtimes}a). The fast relaxation accounts instead for a moderately
viscous response: $\omega (Q)\tau (Q)\sim 1$ around the peak value at $%
Q\simeq 8$ nm$^{-1}$and remains quite lower all around (insets in Fig. \ref
{figtimes}b). Following the arguments of Sec. \ref{memoryfeatures}, we
expect the slow relaxation to be responsible for the narrow quasi elastic
contribution and for the position of the inelastic features of $S(Q,\omega )$
while the fast process should drive mainly the acoustic damping (Brillouin
linewidth) also accounting for the broader part of the Mountain peak \cite
{mountain}.

\begin{figure}[h]
\centering
\includegraphics[width=.8\textwidth]{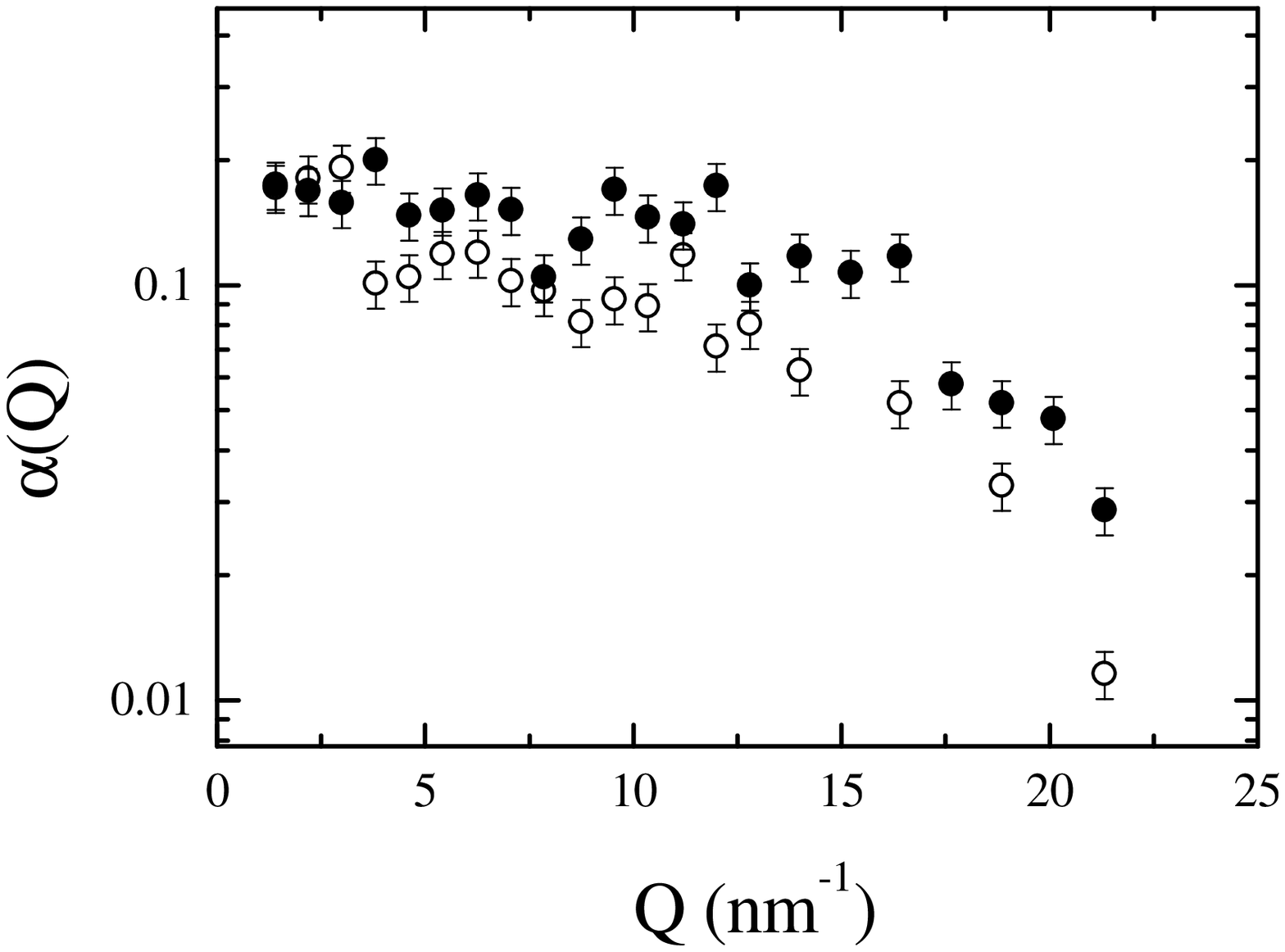}
\vspace{-9cm}
\caption{Ratio between the slow and the total relaxation strength. Full dots
($\bullet $) $T=475$ K, open dots ($\circ $) $T=600$ K.}
\label{figalfaq}
\end{figure}

\begin{figure}[h]
\centering
\includegraphics[width=.8\textwidth]{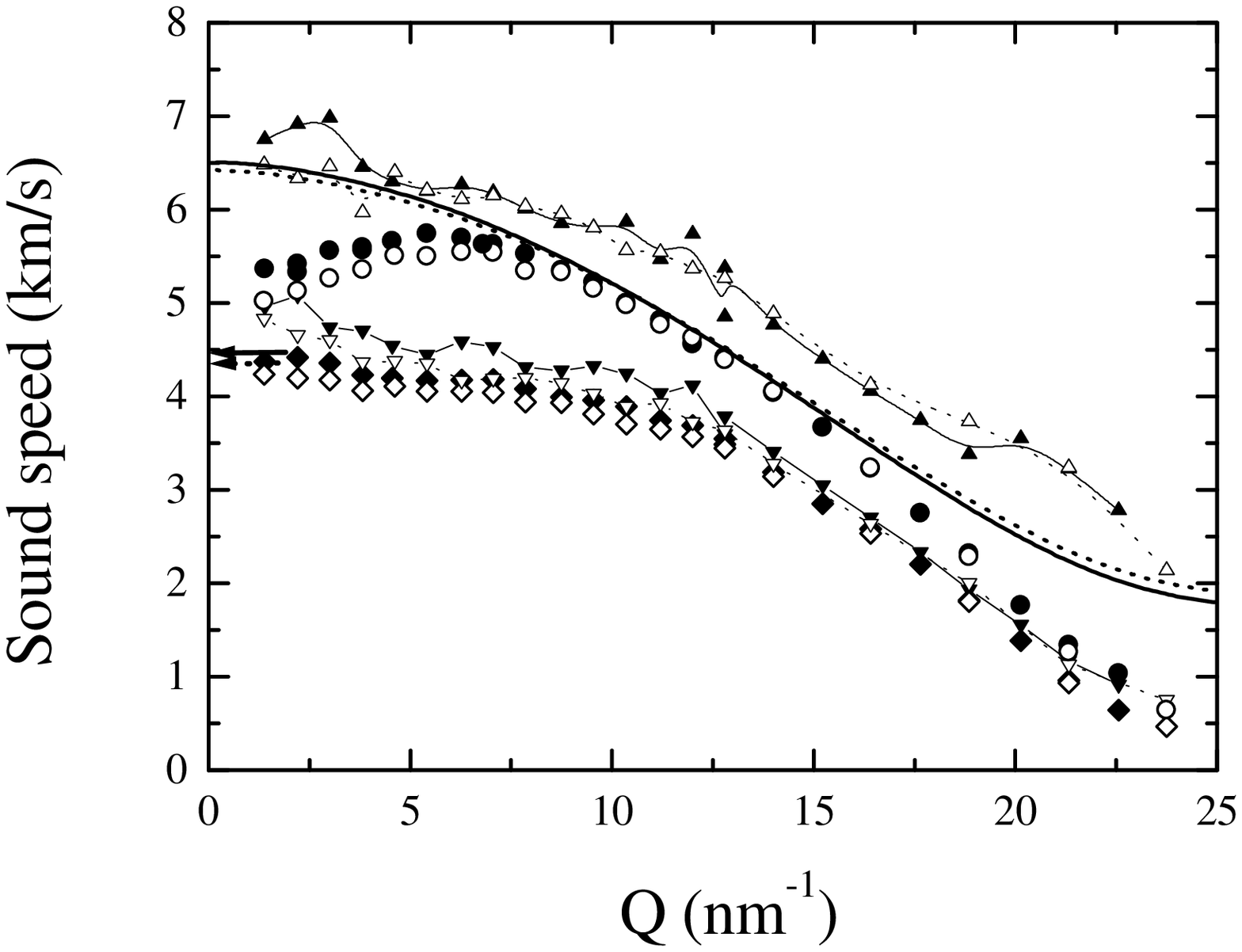}
\vspace{-9cm}
\caption{Apparent sound speed ($\circ $), as from the maxima of $J(Q,\omega
) $ build from the fit parameters (therefore resolution deconvoluted),
compared to the hydrodynamic limit of ref. [33] ($\dashleftarrow $ arrow)
and to the high frequency limit $c_\infty ^{th}(Q)$ ($\cdots \cdots $) of
ref. [28]. The following quantities, deduced from fitting parameters, are
also reported: the isothermal speed of sound $c_0(Q)$ ($\diamondsuit $), the
high frequency limit associated to the slow process $\protect\sqrt{\omega
_0^2(Q)+\Delta _\alpha ^2(Q)}/Q$ ($-\triangledown -$) and finally the total $%
c_\infty (Q)$ ($-\vartriangle -$) with all the strenghts included. The full
lines/symbols denote $T=475$ K data, the dotted lines/full symbols are
relative to $T=600$ K.}
\label{figlowqdisp}
\end{figure}

To go further and point out the real quantitative influence of such two
channels is necessary to look at the strength parameters, or better at the
rate $\alpha (Q).$ Such quantity, defined as the ratio $\Delta _\alpha
^2/\left( \Delta _\mu ^2+\Delta _\alpha ^2\right) ,$ has its maximum of $%
\simeq 0.20$ at low $Q$ and decreases at increasing the wavevector as shown
in Fig. \ref{figalfaq}. This means that the fastest process plays the major
role in characterizing the dynamics of liquid lithium all over the
investigated first Brillouin pseudo-zone. Such statement can be better
understood looking at the apparent sound speed behavior as plotted in Fig.
\ref{figlowqdisp}. The sound velocity moves from the low $Q$ value
(isothermal in the IXS window) to its high frequency value $c_\infty (Q).$
As previously noted, such behavior has been observed in several MD
simulations of alkali metals and in some cases experimentally, even if with
some ambiguity due to kinematic restrictions/incoherent contribution on
neutron side and resolution broadening in the pioneer IXS experiments \cite
{burk}. What is less clear is the microscopic origin of such positive
dispersion. On the basis of our analysis, once unambiguously found out an
experimental evidence of a double timescale, {\it we are able to ascribe the
positive dispersion of the sound velocity to the faster process}. We
reported in fact in Fig. \ref{figlowqdisp} the value of the solid-like
response associated to the single slow process and to both slow and fast
relaxation processes. As expected the ''jump'' in sound velocity associated
to the slow $\alpha $-process ($\left[ \sqrt{\omega _0^2(Q)+\Delta _\alpha
^2(Q)}-\omega _0(Q)\right] /Q$) has already occurred at the lower IXS $Q$
values, but it is quantitatively negligible as compared to the full positive
dispersion of the system. Despite of the fact that the condition $\omega
_l(Q)\tau _\mu (Q)\gg 1$ is never satisfied{\it , the faster process, due to
its dominant strength, accounts almost entirely for the viscous-elastic
transition}. As can be seen in Fig. \ref{figlowqdisp}, the apparent sound
velocity $c_{app}=\omega _l(Q)/Q$, never reaches the infinite frequency
sound velocity deduced by the fit, $c_\infty (Q)=\sqrt{\omega _0^2(Q)+\Delta
_\alpha ^2(Q)+\Delta _\mu ^2(Q)}/Q.$

At first sight this discrepancy may be explained on the fact that $\omega
_l(Q)\tau _\mu (Q)$ is always $\sim 1$. However $c_\infty (Q)$ can be also
evaluated theoretically starting from the knowledge of the interatomic
potential and the pair distribution function (see for example ref. \cite
{libroumberto}). In Fig. \ref{figlowqdisp} we also report the values \cite
{tullioMD} of the unrelaxed sound speed $c_\infty ^{th}(Q)$ computed in this
way, they turns out to be lower than those determined from the fitting of
the spectra at both the temperatures. Moreover, $c_\infty ^{th}(Q)$ is
actually reached by the apparent sound speed at $Q\cong 9$ nm$^{-1},$ i.e.
when $\omega _l(Q)\tau _\mu (Q)\sim 1.$

A possible explanation of the inconsistency between $c_\infty ^{th}(Q)$ and $%
c_\infty (Q)=\sqrt{\omega _0^2(Q)+\Delta _\alpha ^2(Q)+\Delta _\mu ^2(Q)}/Q$
can be related to the arbitrary choice of the memory function details. Even
though the presence of a two timescale dynamics is directly seen in the raw
experimental data (particularly in the $7<Q<14$ nm$^{-1}$ region, the quasi
elastic peak reflects the superposition of two peaks of different widths),
the detailed shape of $M_L(Q,t)$ can have in principle more complicated
features than the double exponential ansatz of Eq. (\ref{x2tempi}). In
particular we already mentioned the major drawback of such an assumption,
namely the cusp at $t=0$. It is reasonable to think that an exponential
decay forced to represent a more complicated time dependence can give an
accurate estimate as far as the $\tau $ is concerned, while at short times
the lack of a zero second derivative inescapably leads to an overestimate of
$M_L(Q,t=0),$ i.e. just the positive dispersion amplitude: a similar effect,
i.e. the overestimation of the $c_\infty (Q\rightarrow 0)$ deduced by the
fit of BLS data using instantaneous approximation for the microscopic
process with respect to the experimental IXS current spectra maxima, has
been recently shown in polybutadiene \cite{fiore}.

\begin{figure}[h]
\centering
\includegraphics[width=.8\textwidth]{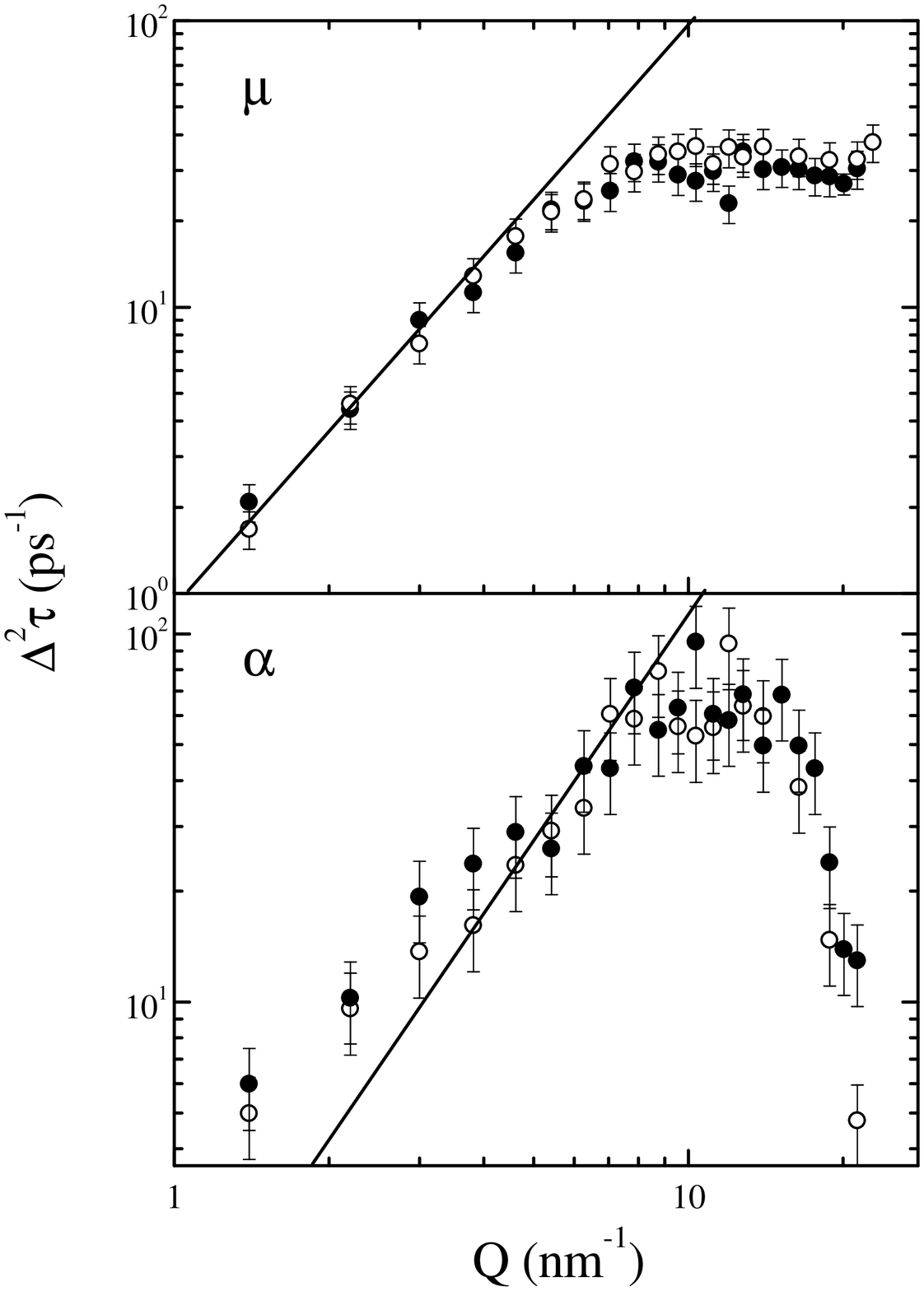}
\caption{Viscosity contribution $\Delta ^2\tau $ of the fast and slow
relaxation, full dots: ($\bullet $) $T=475$ K, open dots ($\circ $) $T=600$
K. $Q^2$ dependence is also shown as comparison (full line (------).}
\label{deltataufig}
\end{figure}

\begin{figure}[h]
\centering
\includegraphics[width=.8\textwidth]{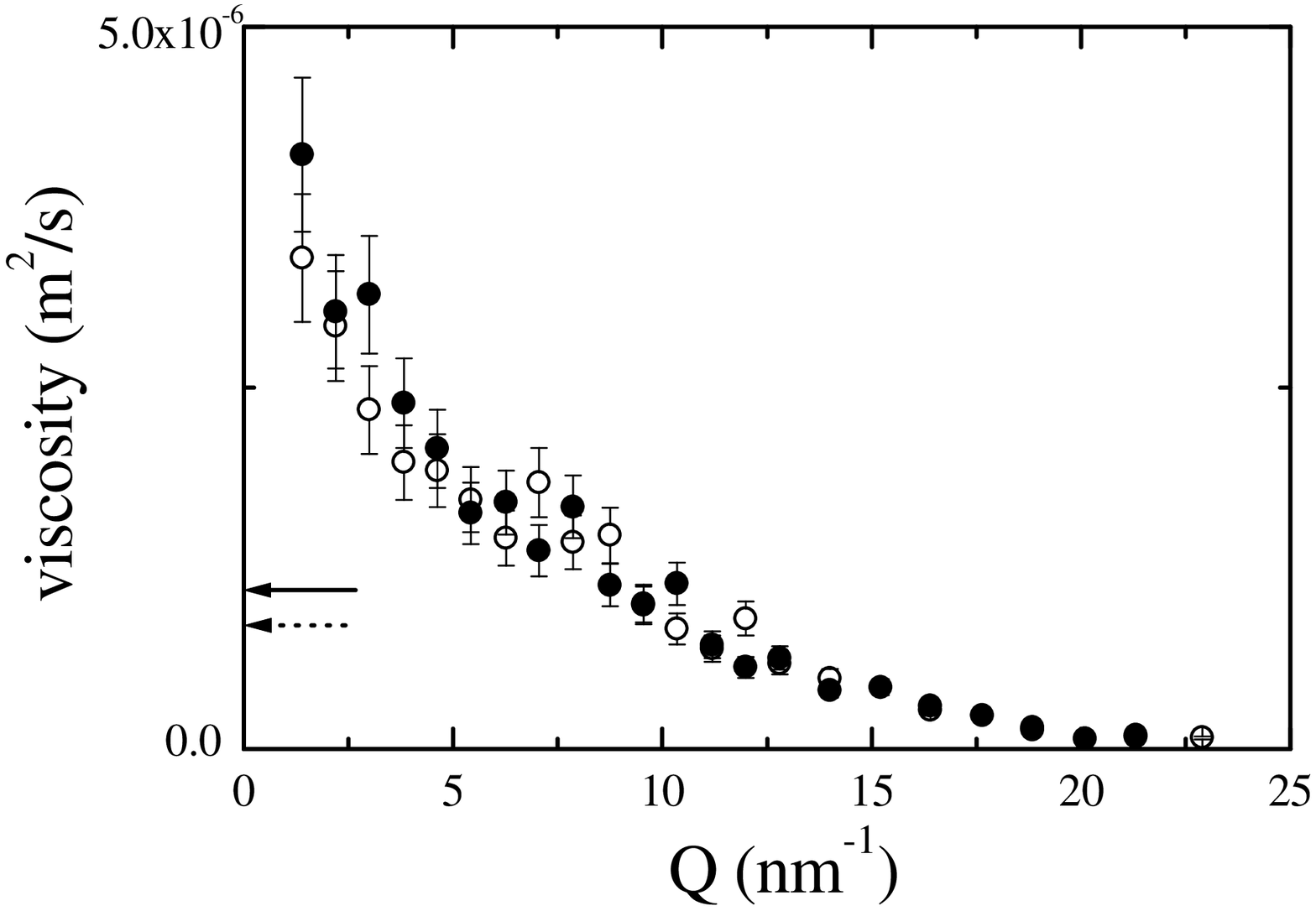}
\vspace{-9cm}
\caption{Longitudinal viscosity from the fit $(\Delta _\alpha ^2\tau _\alpha
+\Delta _\mu ^2\tau _\mu )/Q^2$ compared to shear viscosity data of ref. The
full lines/symbols denote $T=475$ K data, the dotted lines/open symbols are
relative to $T=600$ K.}
\label{figviscosity}
\end{figure}

The last important aspect concerns the relative weights of the two processes
and how to relate them to the acoustic attenuation. The role of the thermal
effects has been already discussed. Looking at the generalized viscosity
contributions $\Delta _\mu ^2\tau _\mu $ and $\Delta _\alpha ^2\tau _\alpha $
(see Fig. \ref{deltataufig}) we find that they are nearly comparable, (the
strength differences are nearly balanced by the different $\tau $), but once
again only the microscopic (faster) relaxation process can be related to the
Brillouin linewidth being the only one close to the $\omega \tau <<1$
condition in the explored $Q$-range. Moreover, in the spirit of the
hydrodynamic generalization, the sum of such contributions is expected to
have a $Q^2$ dependence. Such prediction is quite well reproduced by our
fits: only at very low $Q$ a slight deviation occurs, ascribable to the slow
part; once again the reason may be an excess estimate of the $\alpha -$%
relaxation time at small $Q$. Finally, in Fig. \ref{figviscosity}, the
longitudinal viscosity as outcome of the fit (the full area under $M(Q,t)$)
is compared to an experimental determination of shear static ($Q\rightarrow 0
$) viscosity \cite{ose}. Noting that in molten alkali metals the shear part
usually accounts for a $30-40\%$ of the longitudinal part, the agreement is
satisfactory, except for the very low $Q$ points where finite resolution
function effects prevent an accurate check.

\subsection{The ''high Q region''}

\begin{figure}[h]
\centering
\includegraphics[width=.8\textwidth]{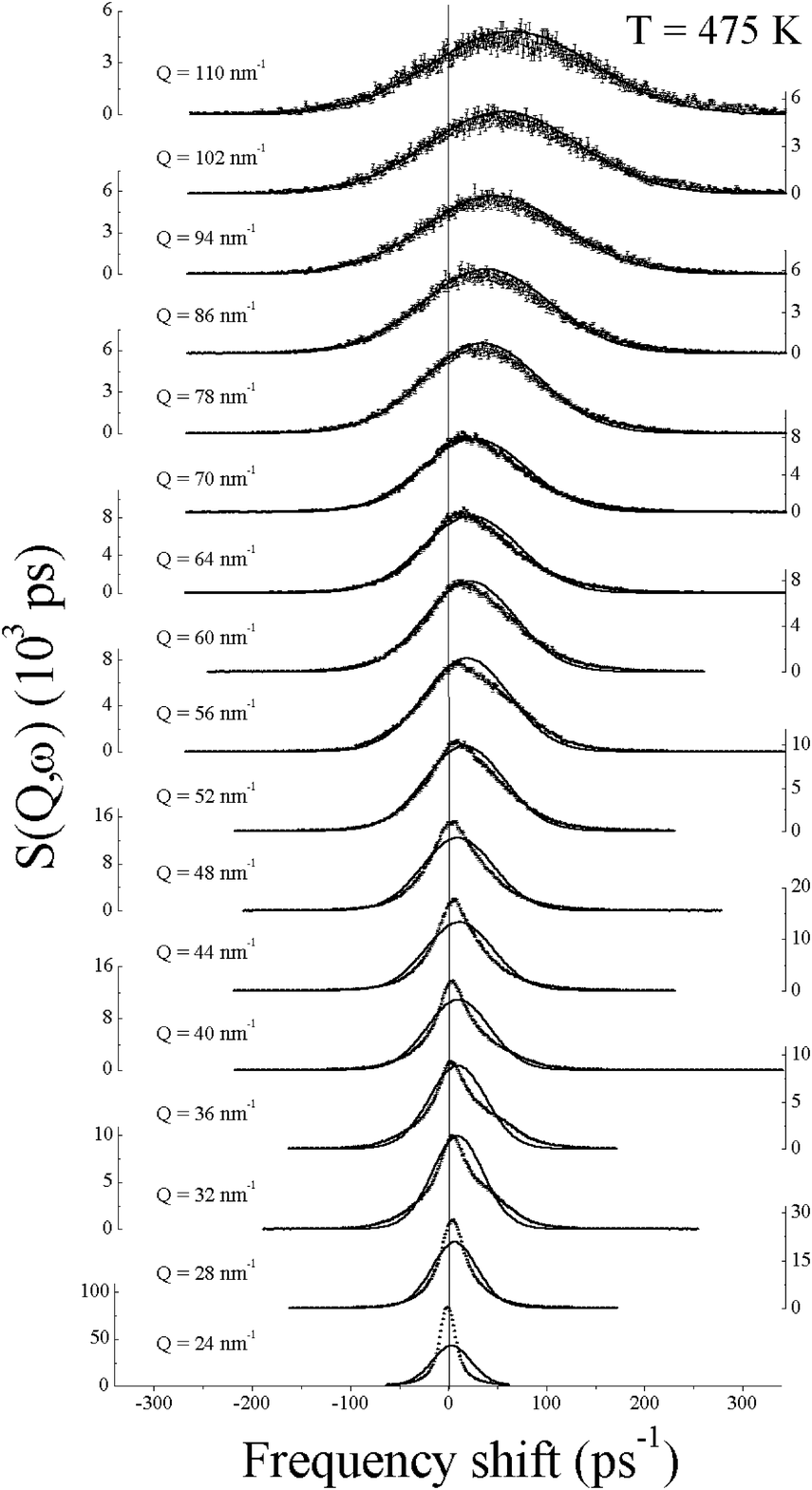}
\caption{IXS spectra at Q values above the main peak of the structure
factor, in the (7 7 7) configuration, at the lower temperature. The full
line is the theoretical lineshape expected in the single particle, high $Q$,
limit.}
\label{IXShighq}
\end{figure}

\begin{figure}[h]
\centering
\includegraphics[width=.8\textwidth]{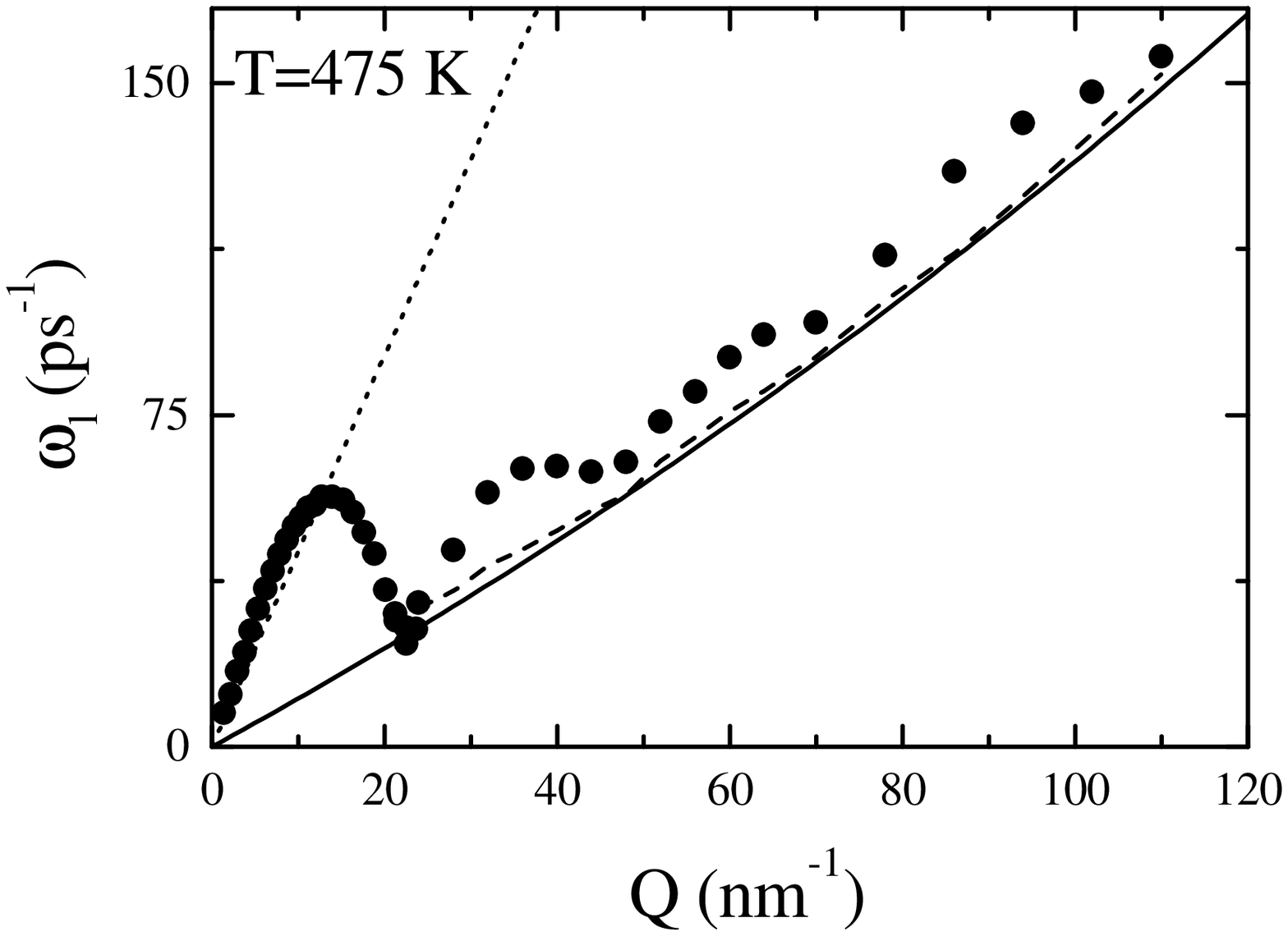}
\vspace{-9cm}
\caption{Maxima of the current spectra ($\bullet $) in the full range of
explored wavevectors at $T=475$ K, compared to the asintotic opposite
behaviors in the low $Q$ (hydrodynamic ($\cdots \cdots $) of ref. [33]) and
high $Q$ (quantum free particle (------), see text) limits. As far as (7 7
7) data are concerned we reported the rough current maxima as calculated
from the spectra. To estimate resolution effect we also include in this case
the high Q limit value as deduced by the free particle model folded with
instrumental resolution.}
\label{fulldisp}
\end{figure}

Above $Q_m$, collective excitations are still evident in $I(Q,\omega )$ for
nearly one more Brillouin pseudo-zone, even if the analysis of the previous
section is not wholly satisfactory because of the worsening of the
experimental resolution. Moreover, single particle features gradually emerge
giving rise to a crossover between two completely different dynamical
regimes. As the density fluctuation wavelength matches the mean
interparticle distance, a transition between strongly correlated to
incoherent motion occurs. The lineshape associated to a quantum free
particle is well known to be Gaussian. Once that $S_q(Q)$ is determined as
before from the first two moments sum rules, there are no unknown
parameters. While in the previous section our sample has been supposed to be
single-component, we must now, due to the emergency of additional quantum
effect, be more accurate by taking into account isotopic effects. From the
natural abundance ratios (92\% of $^7$Li, 8\% of $^6$Li) one finds that

\begin{equation}
S_q(Q,\omega )=\frac 1{\sqrt{2\pi }}\sum_{i=6,7}\frac{C_i}{\sigma _i}e^{{%
-(\omega -\omega _i)^2/2\sigma _i^2}}  \label{sqwhq}
\end{equation}
where $C_i$ are the above mentioned isotope concentrations while
$\omega _i=$ $\hbar Q^2/2m_i$ and ${\sigma _i^2=}K_BTQ^2/m_i$ are
the first (central) and the second (relative to $\omega _i$)
frequency moments of the two isotopes. In Fig. \ref{IXShighq} the
experimental spectra are reported together with the theoretical
lineshapes (resolution convoluted) of Eq. (\ref{sqwhq}) in order
to emphasize all the crossover features. To be more quantitative
we report in Fig. \ref {fulldisp} the Stokes current spectra
maxima $\omega _l$ as function of the exchanged wavevector.
Indeed, in this ``high $Q$'' region (above $Q_m$=24 nm$^{-1}$)
this quantity has been calculated starting from the spectra of
Fig. \ref{IXShighq} (we expect the effect of the resolution
broadening to be negligible at such high $Q's$). For this reason,
due to the quantum features of the $S(Q,\omega)$, the quantity
$\omega_l$ takes different values in the Stokes/Antistokes sides
of the spectra \cite{massimi}. The asymptotic values are expected
to be reached in the very low $Q$ (adiabatic) and high $Q$ (free
particle) \cite{notaquantum}.

After an initial nearly linear dispersion, structural effects take place
suppressing the sound propagation around $Q_{m}/2$ due to strong negative
interference. With increasing $Q$ values, the points in Fig. \ref{fulldisp}
show not only a second pseudo-BZ, but also a series of oscillations that
damp out with increasing $Q$ - here, $\omega _{l}(Q)$ is approaching the
single particle behavior. These oscillations are in anti-phase with those of
$S(Q)$ (see Fig.~\ref{sq}) and can therefore be associated with the local
order in the liquid.

\section{CONCLUSION}

In the present work inelastic X-ray scattering has been utilized to study in
detail the main features of the collective dynamics in liquid lithium, a
system where the use of inelastic neutron scattering is quite difficult and
basically impossible in the low Q region. This IXS experiments have allowed
for a precise assessment of the different relaxation processes which affect
the spectral shape of the dynamic structure factor. Specifically, the
quality of the data required an analysis where one has to invoke the
simultaneous presence of three distinct decay channels of the collective
memory function. The first one is associated with the coupling between
density and temperature fluctuations. As in other alkali metals, although
its relevance is rather small, this process cannot be neglected and plays a
different role in respect to ordinary non-conducting liquids. Much more
important is the unambiguous evidence of {\it two well separated timescales}
in the decay of the $M_L(Q,t)$ - the portion of the memory function directly
associated with generalized viscosity effects. In this respect, our IXS data
prove that the traditional use of a single time-scale (viscoelastic model)
cannot account for the detailed shape of $S(Q,\omega )$, and that it only
provides a guidance for a qualitative description of different dynamical
regimes.

Despite the effect of an additional time-scale to the viscoelastic model has
been tested on MD results in the case of Rubidium (see pag. 248 of ref. \cite
{libroumberto}), and theoretical interpretations of such framework (in
particular devoted to frame the slow process within Mode Coupling theory)
have been recently given \cite{casasMC} we give in this work, to the best of
our knowledge, the first experimental evidence of the {\it necessity} of
such an assumption to correctly reproduce the measured $S(Q,\omega )$,
stressing the role of the contributions from the different relaxations to
the $S(Q,\omega )$ in the considered $Q-E$ regions. The presence of a two
time-scales mechanism leaves open, however, a point of crucial importance:
namely, their physical origin and, most important, of the fast one.

The slow process, referring to the terminology used to describe glass
forming systems, is related to the $\alpha $-relaxation responsible for the
liquid-glass transition in systems capable to sustain strong supercooling.
The same process can be framed within kinetic theory in terms of
''correlated collisions'': in mode coupling approaches, the onset of these
correlation effects is traced back to the coupling to slowly relaxing
dynamical variables, and specifically in the liquid region, to long lasting
density fluctuations. The corresponding decay mechanism (often referred to
as ''structural relaxation''), almost negligible in the initial collisional
region, controls the dynamics at intermediate and long times. For $%
Q\rightarrow 0$ the amplitude of the slow portion of the memory function
turns out to vanish as $Q^2$, whereas at increasing wavevectors it becomes
proportional to $Q$ \cite{beng}. At sufficiently large $Q$, the relevance of
these couplings as long lasting decay channel decreases markedly. In fact,
in a rather wide portion of the explored range of wavevectors, we find that
the amplitude $\Delta _L^2(Q)$ of the {\it total} memory function behaves
approximately as $Q^2.$ Therefore, outside the hydrodynamic region, the
normalized weight of the slowly varying contributions to $M_L(Q,t)$
approximately decreases proportionally to $Q/Q^2=1/Q.$ This weight may
reasonably be identified with the dimensionless amplitude factor $\alpha (Q)$
in the model of Eq. (\ref{x2tempi}). A quantitative understanding of the
''slow'' timescale would require a full evaluation of the mode-coupling
contribution at different wavevectors, as in fact done in the above
mentioned MD study \cite{casasMC}.

The situation is much more involved as far as the fast process is concerned
and, in fact, its microscopic interpretation is still matter of debate. In
the present work we have given an experimental evidence that, in liquid Li,
the fast process is indeed the dominant one, and {\it it mainly controls
both the sound speed dispersion and attenuation} i.e. the position and the
width of the Brillouin component of the spectra. This observation justifies
the need for a detailed discussion on the nature of the fast relaxation
channel. Within generalized kinetic theory the fast initial decay of both
single particle and collective memory function is traced back to collisional
events, which are short ranged both in space and in time. In particular, in
the collective case, the short timescale $1/\gamma _1(Q)$ can be associated
with the duration of a rapid structural rearrangement occurring over a
spatial range $\cong 2\pi /Q_m.$ In most liquids, at low and intermediate
wavevectors, the typical value of the time $1/\gamma _1(Q)$ is of the order
of a few $10^{-14}$ s, with some structural oscillations of decreasing trend
\cite{sjogr}. At $Q>Q_m$, the structural effects damp out, the decrease
becomes more marked, and, when the single-particle aspects start to prevail,
$1/\gamma _1(Q)\propto Q^{-1}.$ An inspection of the results reported in a
recent MD study in liquid lithium near melting \cite{casasMC} basically
confirms these qualitative features as well as the actual values of the
''short time'' found in our data analysis (see Fig. \ref{figtimes}).
Although the description of the fast process in term of interparticle
collision is a possible way to account for the dynamical features at short
times, it does not allow a deep insight into the physics behind it.

Although the description of the fast process in terms of local collisional
events has been widely used in the past to account for the short-time
dynamics, other mechanisms cannot be excluded. We have already mentioned in
sec. II that a formally similar ''microscopic mechanism'' has been
introduced in the analysis of the Brillouin spectra of glassformers \cite
{cumm}; more recently, an analogous microscopic relaxation process has been
detected in a simulated harmonic glass \cite{glasslett}. Generally speaking,
the inclusion of a relaxation process in a memory function reflects the
existence of decay channels by which the energy stored in a given degree of
freedom (''mode'') relaxes toward other modes. The problem is then the
physical identification of the modes entering the fast relaxation process,
and the ultimate origin of the latter in a widely different class of systems
ranging from fluids to glasses. In generalized kinetic theories the
mechanism is traced back to a mainfold of non-hydrodynamic, phase-space
''kinetic'' modes, whose non-conserved character ensures a rapid decay of
the memory function \cite{sjogr,desh}. A different approach to describe the
fast part of the memory function in disordered systems relies on the normal
modes (''instantaneous'' in normal liquids) analysis of the atomic dynamics
\cite{keyes}. In this case one uses a framework (dynamical matrix, etc.)
formally similar to the one customarily adopted for harmonic crystals;
however, owing to the lack of translational symmetry of the system, it turns
out \cite{glasslett} that the eigenstates cannot anymore be represented by
planewaves (PW) even at relatively small wavevectors \cite{mazza}. As a
result, when probing the system at a specific wavevector $Q$ one effectively
detects a transfer from a PW mode toward other PWs of different wavevectors.
Assuming that it is actually this energy flow which causes the fast
relaxation, we deal with a mechanism whose ultimate origin is the
topological disorder, and not a truly dynamical event. Ordinary liquids
(such as the one considered here) are certainly ''disordered systems'' and
at the high frequency considered here ($\omega \tau _\alpha >>1$), they can
be considered as ''frozen''; thereforethe description used for glasses can
be applied as well. At the present stage of the theory our experimental data
cannot support either interpretation, or not even ascertain whether the
distinction is largely semanthic, or not. However, the posssibility that the
phenomenology reported here for a simple liquid can be understood within the
same mental framework developed for more complex liquids and glasses, is
certainly interesting. Further studies on other simple fluids, as for
example neon \cite{neon}, seem to suggest that is indeed the case.

\section{ACKNOWLEDGMENTS}
We gratefully aknowledge the ID-16 (ESRF) beamline staff for the assistence during the experiments preparation and performance. One of us (T. Scopigno) feels to be indebted with Dr. R. Di Leonardo for many fruitful discussions.


\onecolumn
``Density Fluctuations in Molten Lithium...'' by T.Scopigno et al.


\newpage
``Density Fluctuations in Molten Lithium...'' by T.Scopigno et al.

\newpage
``Density Fluctuations in Molten Lithium...'' by T.Scopigno et al.

\newpage
``Density Fluctuations in Molten Lithium...'' by T.Scopigno et al.

\newpage
``Density Fluctuations in Molten Lithium...'' by T.Scopigno et al.

\newpage
``Density Fluctuations in Molten Lithium...'' by T.Scopigno et al.

\newpage
``Density Fluctuations in Molten Lithium...'' by T.Scopigno et al.

\newpage
``Density Fluctuations in Molten Lithium...'' by T.Scopigno et al.

\newpage
``Density Fluctuations in Molten Lithium...'' by T.Scopigno et al.

\newpage
``Density Fluctuations in Molten Lithium...'' by T.Scopigno et al.

\newpage
``Density Fluctuations in Molten Lithium...'' by T.Scopigno et al.

\newpage
``Density Fluctuations in Molten Lithium...'' by T.Scopigno et al.


\begin{references}
\bibitem{notasq}  In this respect it is worth to point out as the $S(Q)$
calculated within classical the approximation (MD) may in principle be
different from the real one (IXS). Actually it has already been shown that
an excellent agreement exists between MD and INS experiments \cite
{umblitiomd} so that such a comparison is fully justified.
\bibitem{massimi}  Below $Q_m$ instead, we calculated $\omega_l$ from the
theorethical classical model adjusted to the experimental data, so that it
has the same values in the Stokes/Antistokes side of the spectra.
\bibitem{notaquantum}  For a single component system such quantities,
different on the Stokes/Antistokes side, can be calculated analytically by
solving $\partial \left[ \omega ^2S(Q,\omega )\right] /\partial \omega =0.$
It turns out $\omega _l(Q)=\left( \omega _0\pm \sqrt{\omega _0^2+8\sigma ^2}%
\right) /2;$ we numerically solved this equation for the isotopic mixture.
\end{references}
\end{document}